%% file: main.tex
\documentclass[10pt,twocolumn,letterpaper]{article}
\usepackage{authblk}
\usepackage[pagenumbers]{cvpr} 


\usepackage{graphicx}
\usepackage{epstopdf}
\usepackage{bbding}
\usepackage{nicematrix,booktabs,caption}
\usepackage{multirow}
\usepackage{CJKutf8}
\usepackage{latexsym}
\definecolor{iccvblue}{rgb}{0.21,0.49,0.74}
\usepackage[pagebackref,breaklinks,colorlinks,allcolors=iccvblue]{hyperref} 

\definecolor{babyblue}{rgb}{0.54, 0.81, 0.94}

\title{HarmonySeg: Tubular Structure Segmentation with \\Deep-Shallow Feature Fusion and Growth-Suppression Balanced Loss}

\author[1,2,3]{Yi~Huang\thanks{Both authors contributed equally to this work.}}
\author[4]{Ke~Zhang$^*$}
\author[1,2]{Wei~Liu}
\author[3]{Yuanyuan~Wang}
\author[4]{Vishal~M.~Patel}
\author[1]{Le~Lu}
\author[5]{Xu~Han}
\author[1]{Dakai~Jin}
\author[1,2]{Ke~Yan\thanks{Corresponding author, yanke.yan@alibaba-inc.com}}
\affil[1]{DAMO Academy, Alibaba Group}
\affil[2]{Hupan Lab, Hangzhou, China}
\affil[3]{Department of Biomedical Engineering, Fudan University}
\affil[4]{Department of Electrical and Computer Engineering, Johns Hopkins University}
\affil[5]{Department of Hepatobiliary and Pancreatic Surgery, The First Affiliated Hospital of College of Medicine, Zhejiang University}

\begin{document}
\begin{CJK}{UTF8}{gbsn}
\maketitle
\input{sec/0_abstract}    
\input{sec/1_intro}
\input{sec/2_related_work}
\input{sec/3_method}

\input{sec/4_experiment}
\input{sec/5_conclusion}
{
    \small
    \bibliographystyle{ieeenat_fullname}
    \bibliography{main}
}

\end{CJK}
\end{document}


\twocolumn[
\centering
\Large
\textbf{HarmonySeg: Tubular Structure Segmentation with \\Deep-Shallow Feature Fusion and Growth-Suppression Balanced Loss} \\
\vspace{0.5em}Supplementary Material \\
\vspace{1.0em}
]
In the supplementary material, we provide detailed explanations of the vesselness filter (Appendix~\ref{sec:vessel}), the flexible convolution block (Appendix~\ref{sec:convolution}), the segmentation fusion in D2SD (Appendix~\ref{sec:segfusion}), along with additional experimental and visualization results, including refined hepatic vessel labels (Appendix~\ref{sec:refine}), ablation studies (Appendix~\ref{sec:ablation}), our curated HVS-External dataset (Appendix~\ref{sec:hvs}), the robustness of loss functions (Appendix~\ref{sec:robust}), and the trade-off between precision and recall of loss functions (Appendix~\ref{sec:tradeoff}).
\appendix
%

\begin{figure}[t]
\centering
\includegraphics[width=\linewidth]{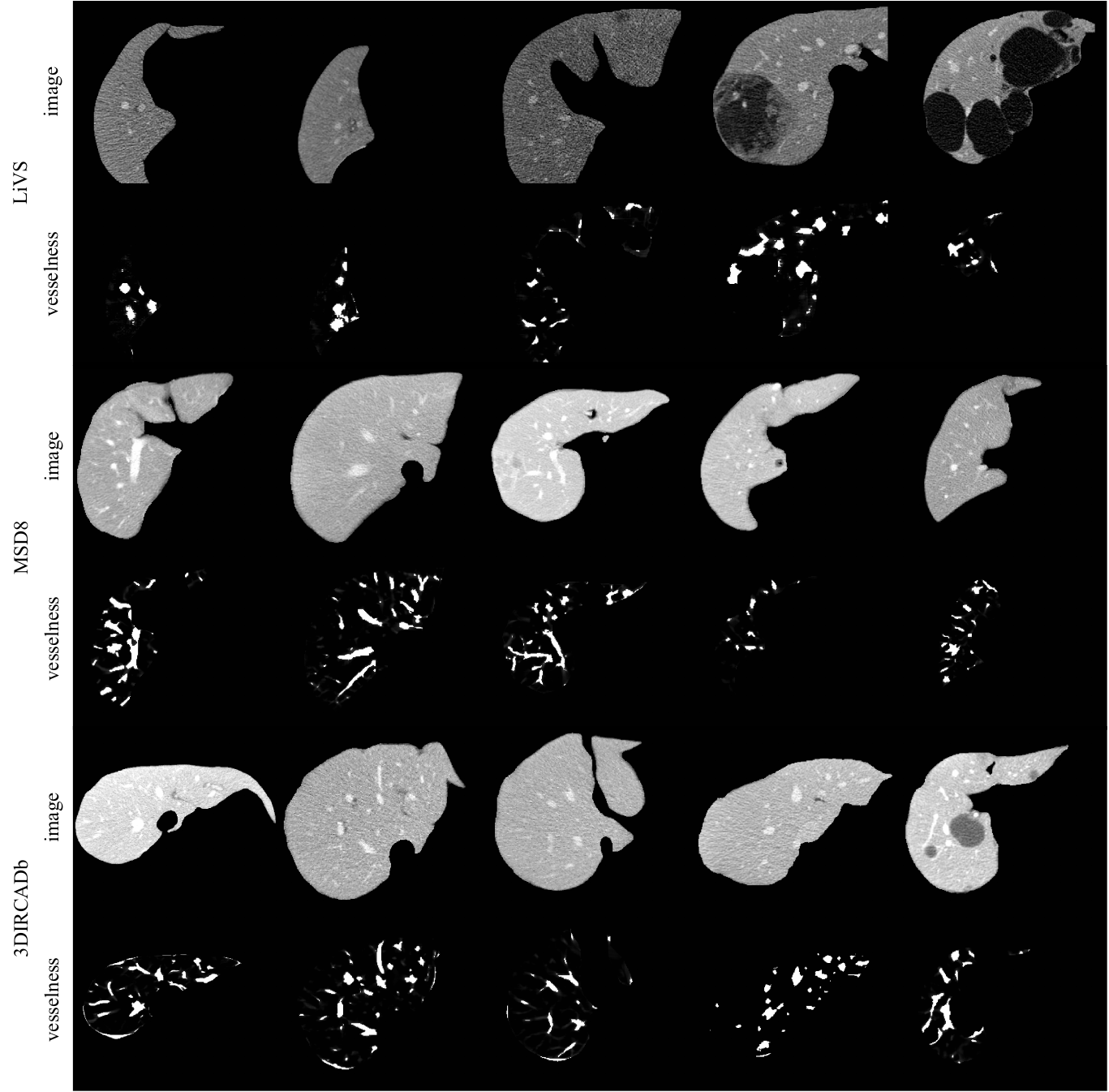}
\caption{Visualization of images and their corresponding vesselness filtering results. In the CT images, regions with high intensities represent hepatic vessels, while dark regions indicate tumors. In the paired vesselness filtering results, high-intensity patches correspond to vessel candidates, with noise visibly present along the liver border.}
\label{fig:Vesselness_visual}
\end{figure}

\begin{table}[]
\centering
\caption{Enhanced ratio of labeled slices on the HVS task brought by refined labels (\%).}\label{table:LabeledRatios}
\begin{tabular}{ccclc}
\hline
\multicolumn{5}{c}{Hepatic vessel segmentation (HVS)}                                         \\ \hline
\multicolumn{1}{c|}{Label} & LiVS        & \multicolumn{2}{c}{MSD8}        & \begin{tabular}[c]{@{}c@{}}3DIRCADb\\ (Testing)\end{tabular} \\ \hline
\multicolumn{1}{c|}{Original} & 20.26\textsubscript{±12.77} & \multicolumn{2}{c}{71.96\textsubscript{±12.29}} & \multirow{2}{*}{78.87\textsubscript{±7.27}} \\
\multicolumn{1}{c|}{Refined}       & 70.26\textsubscript{±11.75} & \multicolumn{2}{c}{72.54\textsubscript{±11.90}} &          \\ \hline
\end{tabular}
\end{table}


\begin{table}[]
\centering
\caption{Quantitative improvement on the HVS task brought by refined labels. The segmentation model is nnU-Net.}\label{table:LabelsComparison}
\resizebox{\linewidth}{!}{
\begin{tabular}{cccc}
\hline
\multirow{2}{*}{Dataset} & \multirow{2}{*}{Label} & \multirow{2}{*}{Dice(\%,↑)} & \multirow{2}{*}{HD(↓)} \\
                                                                    &                               &                             &                        \\ \hline
\multirow{2}{*}{HVS}  & Original         & 56.69\textsubscript{±8.42}                            & 10.83\textsubscript{±4.28}                         \\
  &  Refined                             & \textbf{60.15}\textsubscript{±9.49}                  & \textbf{10.00}\textsubscript{±4.22}             \\ \hline
\multirow{2}{*}{HVS-External}  & Original & 63.09\textsubscript{±10.90}          & 4.02\textsubscript{±1.43}                     \\
  & Refined                              & \textbf{63.78}\textsubscript{±9.27}                  & \textbf{3.69}\textsubscript{±1.17}              \\ \hline
\end{tabular}}
\end{table}

\begin{table*}[]
\centering
\caption{Quantitative comparison on the HVS-External test set stratified by diseases of the subjects.}\label{table:Further_HVSExternal}
\resizebox{\textwidth}{!}{
\begin{tabular}{ccccccccc}
\hline
\multicolumn{9}{c}{HVS-External}                                                                                                                   \\ \hline
\multirow{2}{*}{Model} & \multicolumn{2}{c}{Fatty liver} & \multicolumn{2}{c}{Cirrhosis} & \multicolumn{2}{c}{Tumor} & \multicolumn{2}{c}{Healthy} \\
                       & Dice(\%,↑)      & HD(↓)         & Dice(\%,↑)     & HD(↓)        & Dice(\%,↑)   & HD(↓)      & Dice(\%,↑)    & HD(↓)       \\ \hline
nnU-Net{~\cite{isensee2021nnu}}                & 53.97\textsubscript{±1.97}      & 3.90\textsubscript{±0.09}     & 69.83\textsubscript{±9.15}     & 2.78\textsubscript{±0.64}    & 62.88\textsubscript{±10.14}  & 3.91\textsubscript{±1.29}  & 62.09\textsubscript{±2.75}    & 3.94\textsubscript{±1.08} \\
nnU-Net w clDiceLoss{~\cite{shit2021cldice}}   & 50.80\textsubscript{±3.10}      & 4.16\textsubscript{±0.11}     & 63.48\textsubscript{±2.72}     & 3.51\textsubscript{±0.18}    & 57.18\textsubscript{±7.33}   & 4.40\textsubscript{±1.31}  & 49.61\textsubscript{±8.00}    & 4.96\textsubscript{±0.72}   \\
nnU-Net w SRLoss{~\cite{kirchhoff2024skeleton}}       & 58.88\textsubscript{±1.69}      & 4.08\textsubscript{±1.23}     &\textbf{81.05}\textsubscript{±2.22}     & 3.21\textsubscript{±0.92}    & 70.28\textsubscript{±9.60}   &\underline{3.35}\textsubscript{±1.16}  & 68.73\textsubscript{±4.72}    &\underline{3.56}\textsubscript{±1.16}   \\
TransU-Net3D{~\cite{chen2024transunet}}            & 51.68\textsubscript{±7.78}      & 6.03\textsubscript{±2.99}     & 38.18\textsubscript{±31.76}    & 4.43\textsubscript{±1.73}    & 39.93\textsubscript{±21.86}  & 9.76\textsubscript{±9.17}  & 56.60\textsubscript{±12.92}   & 5.23\textsubscript{±1.67}   \\
DSC-Net{~\cite{qi2023dynamic}}                &\underline{71.44}\textsubscript{±4.51}      &\textbf{3.10}\textsubscript{±0.39}     & 80.50\textsubscript{±3.68}     &\underline{2.73}\textsubscript{±0.66}    &\underline{74.86}\textsubscript{±7.04}   &\textbf{3.18}\textsubscript{±1.34}  &\underline{71.26}\textsubscript{±7.90}    & 3.69\textsubscript{±1.10}   \\
HarmonySeg             &\textbf{73.15}\textsubscript{±1.83}      &\underline{3.74}\textsubscript{±0.11}     &\underline{80.59}\textsubscript{±4.61}     &\textbf{2.20}\textsubscript{±0.55}    &\textbf{75.67}\textsubscript{±7.36}   & 3.61\textsubscript{±1.31}  &\textbf{74.75}\textsubscript{±6.05}    &\textbf{3.37}\textsubscript{±1.18}   \\ \hline
\end{tabular}}
\end{table*}

\section{Vesselness Filter}
\label{sec:vessel}
In this section, we elaborate on the details of vesselness filters and present several vesselness maps for visualization. Image derivatives, including first-order derivatives for border detection and second-order derivatives for shape extraction, are commonly used to highlight vascular structures in images~\cite{agam2005vessel}. Hessian matrix analysis is a representative method based on second-order derivates that can distinguish rounded, tubular, and planar structures~\cite{frangi1998multiscale}. Vesselness filters also employ eigen-decomposition of the Hessian matrix to measure tubularity and enhance vessel regions~\cite{lamy2022benchmark}. Let $\mathit{H}$ be the hessian matrix of a voxel in CT volume, and $\mathbf{e_1}$, $\mathbf{e_2}$ and $\mathbf{e_3}$ be the three eigenvectors of $\mathit{H}$ with corresponding eigenvalues of $\mathit{\lambda}_1$, $\mathit{\lambda}_2$ and $\mathit{\lambda}_3$ ($|\mathit{\lambda}_1|\leq|\mathit{\lambda}_2|\leq|\mathit{\lambda}_3|$). The tubularity is defined as~\cite{lorenz1997multi}:

\setlength{\belowdisplayskip}{0pt} \setlength{\belowdisplayshortskip}{0pt}
\setlength{\abovedisplayskip}{0pt} \setlength{\abovedisplayshortskip}{0pt}
\begin{equation}
|\mathit{\lambda}_1|\approx0,\mathit{\lambda}_2\approx\mathit{\lambda}_3\ll0.
\end{equation}

\begin{figure}[!t]
\centering
\includegraphics[width=0.6\linewidth]{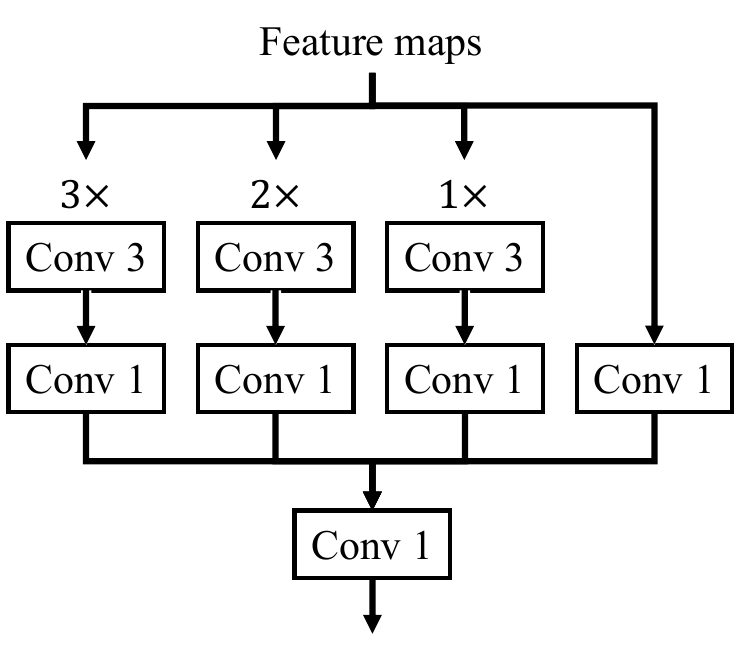}
\caption{Details of flexible convolution block with diverse receptive fields.}
\label{fig:flexibleConv}
\end{figure}

\begin{figure}[t]
\centering
\includegraphics[width=\linewidth]{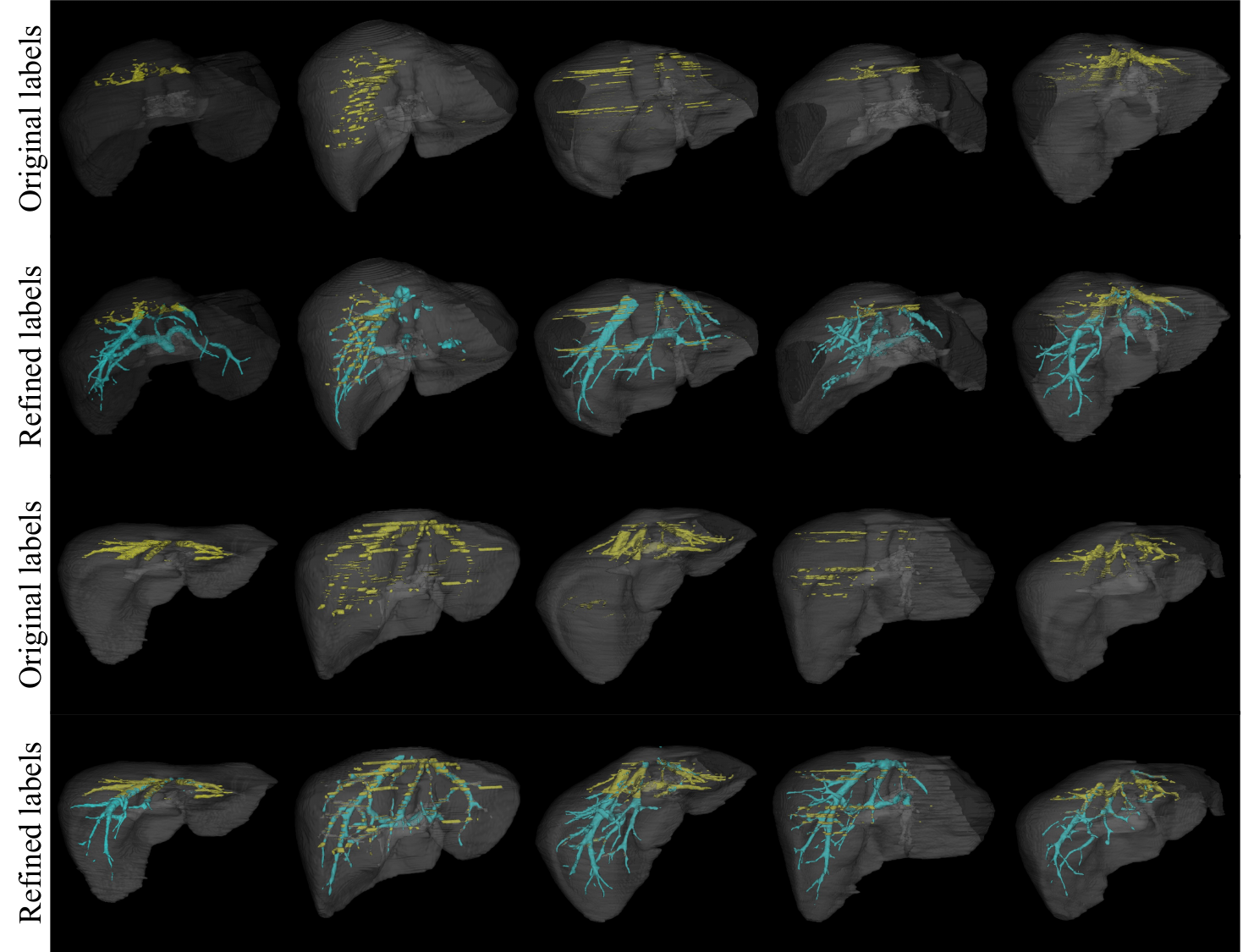}
\caption{Visualization of original and refined hepatic vessel labels in LiVS. The liver is rendered in gray, while the original and refined vessel labels are denoted in yellow and cyan, respectively.}
\label{fig:RefinedLabels_LiVS}
\end{figure}

\begin{figure}[t]
\centering
\includegraphics[width=\linewidth]{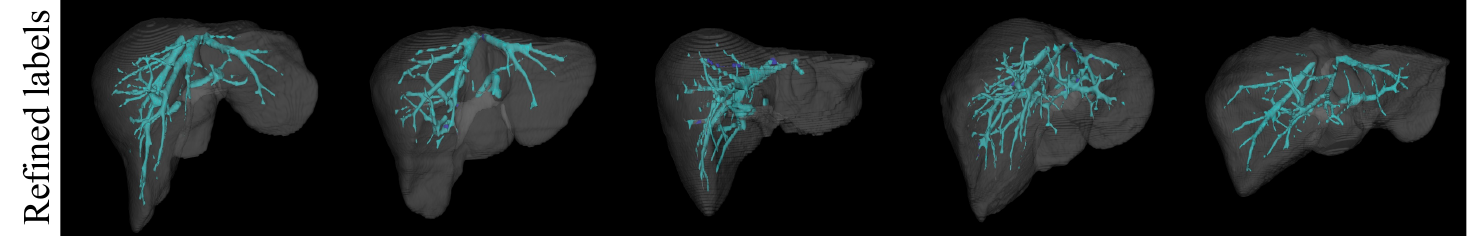}
\caption{Visualization of predicted hepatic vessel labels in the test set of MSD8. The gray color shows the liver and the cyan color denotes the labels. Note that the original label of MSD8's test set is unavailable.}
\label{fig:RefinedLabels_MSD8}
\end{figure}

\begin{figure*}[!t]
\centering
\includegraphics[width=0.65\textwidth]{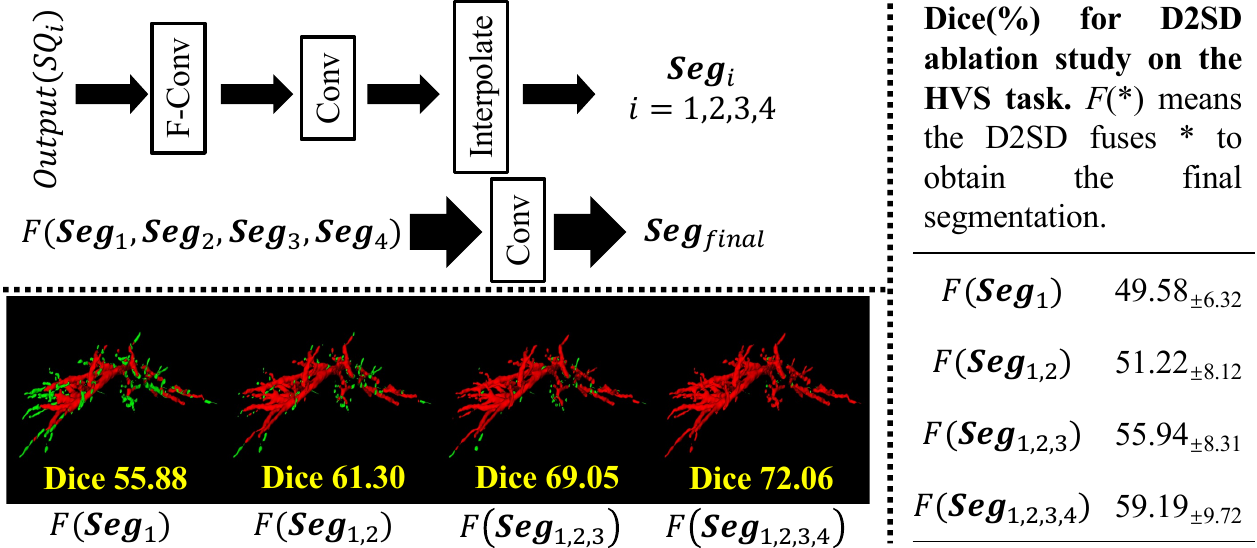}
\caption{Qualitative and quantitative analysis of segmentation fusion in D2SD: red/green indicating the segmentation and the corresponding labels.}
\label{fig:d2sd}
\end{figure*}

Based on this, the Jerman~\cite{jerman2016enhancement} vesselness filter used in our framework further regularizes $\mathit{\lambda}_3$ to reduce the sensitivity for those low-contrast regions:
\begin{equation}
\mathit{F} =
\begin{cases}
0,       &\mathit{\lambda}_2\leq0 \text{ or } \mathit{\lambda}_p\leq0,\\
1,   &\mathit{\lambda}_2\geq\frac{\mathit{\lambda}_p}{2}>0,\\
\mathit{\lambda}^2_2(\mathit{\lambda}_p-\mathit{\lambda}_2)(\frac{3}{\mathit{\lambda}_p+\mathit{\lambda}_2})^3,           & \text{otherwise},\\
\end{cases}
\end{equation}
in which:
\begin{equation}
\mathit{\lambda}_p =
\begin{cases}
\mathit{\lambda}_3, &\mathit{\lambda}_3> \tau \text{ }\text{max}_x\text{ } \mathit{\lambda}_3(x),\\
\mathit{\lambda}_3> \tau \text{ }\text{max}_x\text{ } \mathit{\lambda}_3(x),   &0<\mathit{\lambda}_3 \leq \tau \text{ }\text{max}_x\text{ } \mathit{\lambda}_3(x),\\
0, & \text{otherwise},\\
\end{cases}
\end{equation}
where $\tau\in[0,1]$. Benefiting from this regularization, the Jerman vesselness filter becomes robust even when facing non-homogeneous vessel intensity.
To reveal the effectiveness of the vesselness filter, we give examples of paired vesselness filtering results in Figure~\ref{fig:Vesselness_visual}. As shown in the figure, the vesselness filter can highlight liver vessel candidates of different sizes, even for cases in which tumors exist.
\begin{figure}[t]
\centering
\includegraphics[width=\linewidth]{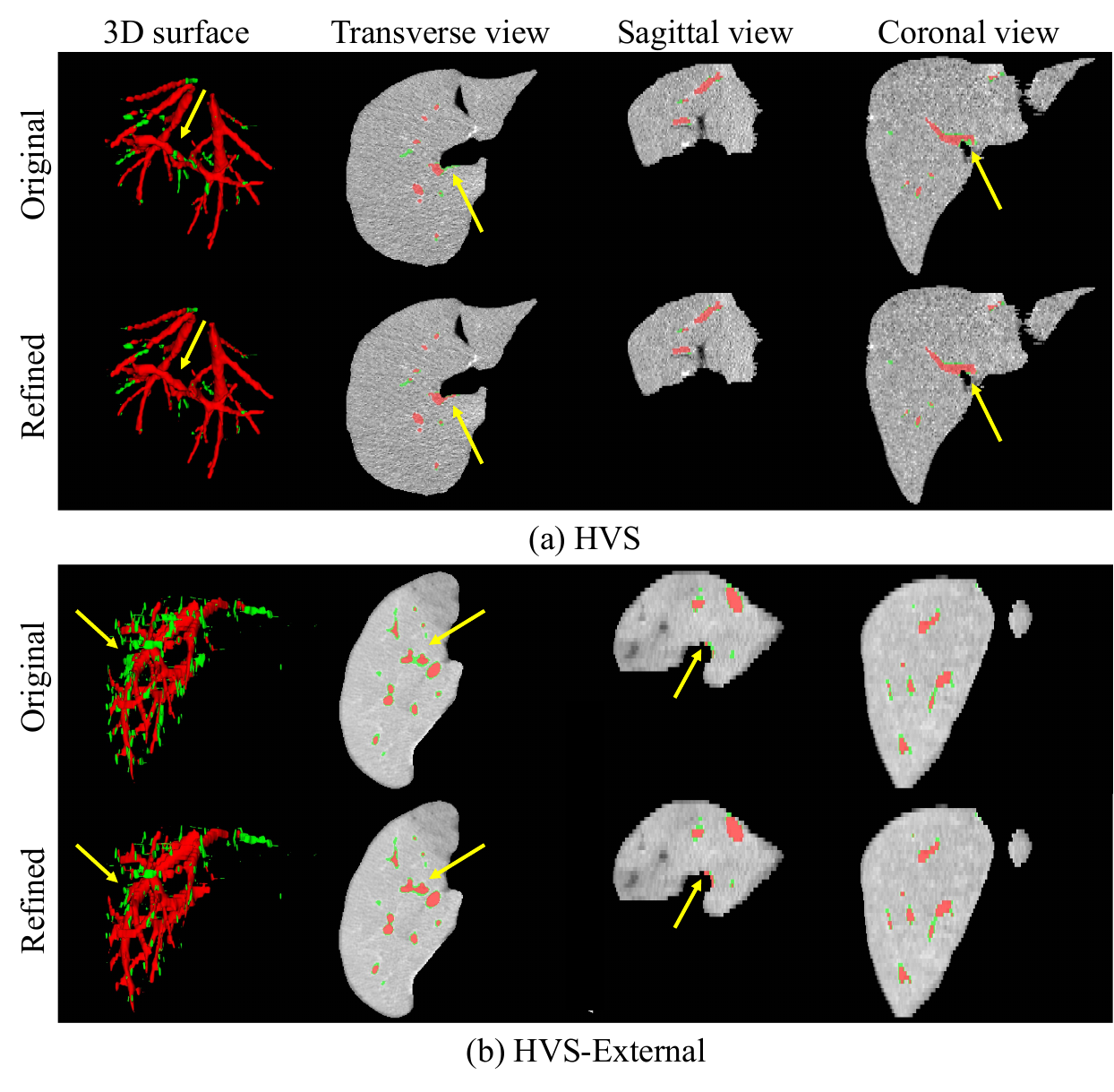}
\caption{Visualization of hepatic vessel segmentation results using nnU-Net trained on both original and refined labels, in which the red indicates the segmentation and the green indicates the corresponding labels. Improvements are highlighted with yellow arrows.}
\label{fig:originalvsrefined}
\end{figure}

\begin{figure}[t]
\centering
\includegraphics[width=\linewidth]{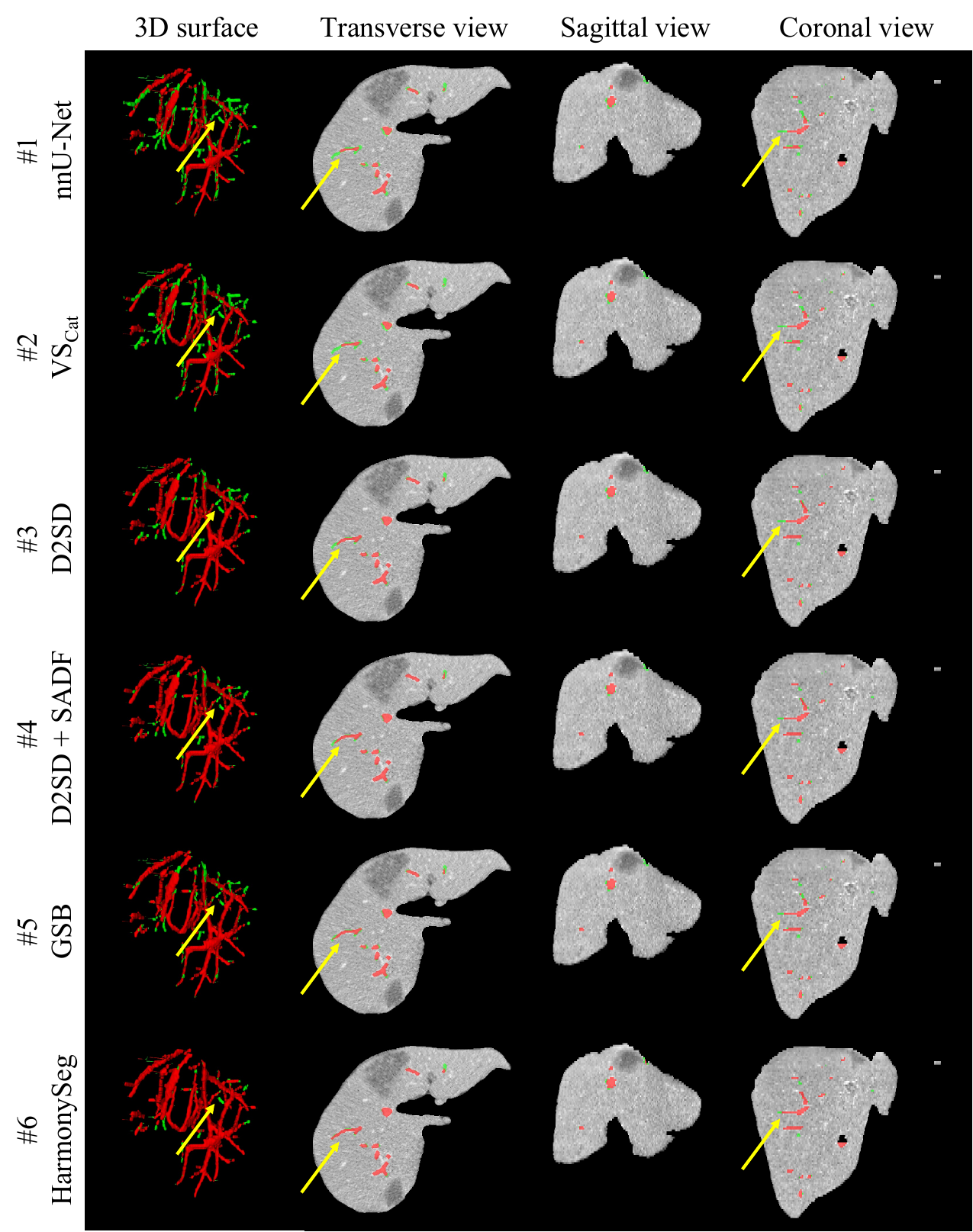}
\caption{Visualization of hepatic vessel segmentation results in ablation study, with red indicating the segmentation and green representing the corresponding labels. Improvements are highlighted with yellow arrows.}
\label{fig:HVS_abaltion}
\end{figure}

\begin{figure}[t]
\centering
\includegraphics[width=\linewidth]{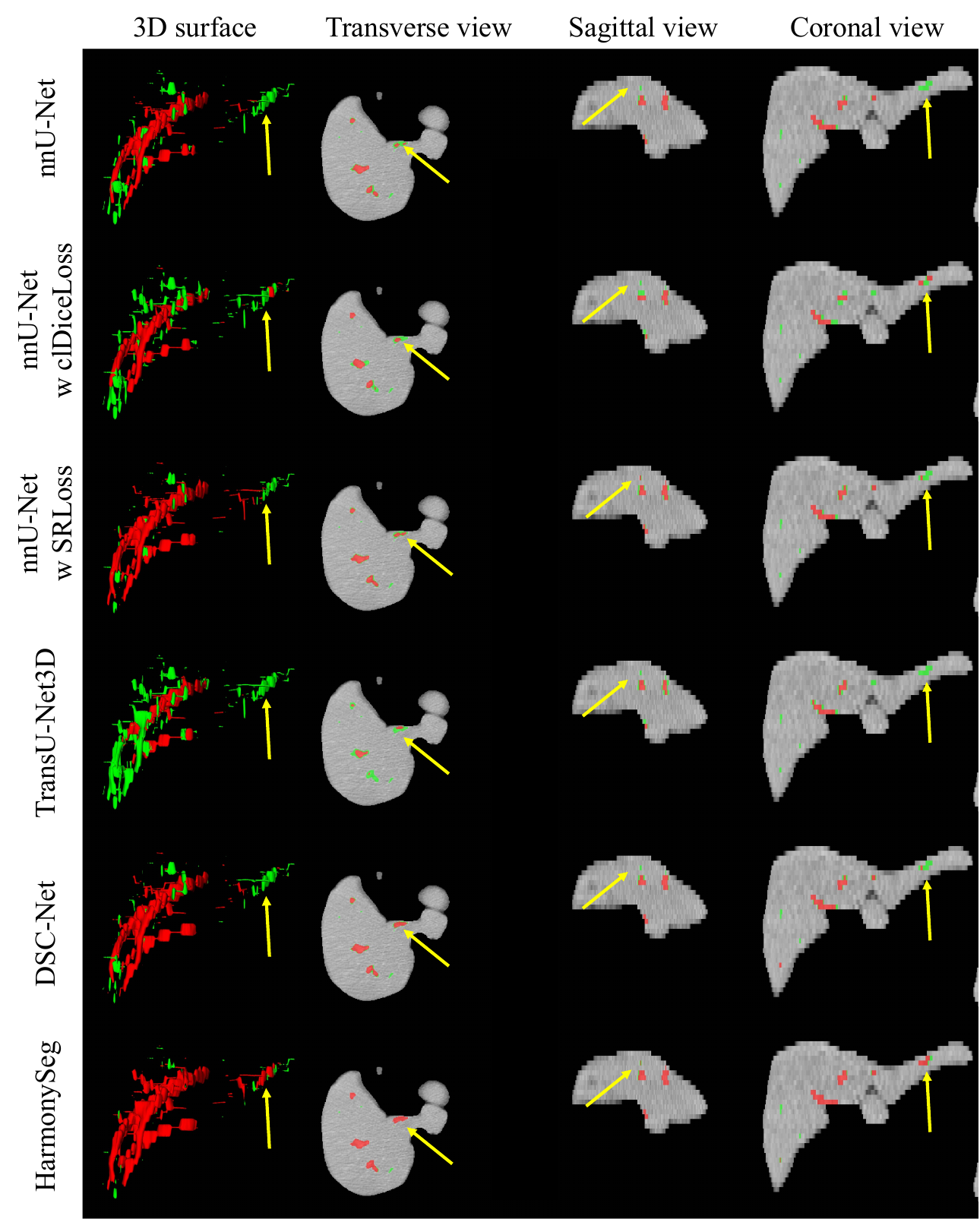}
\caption{Visualization of hepatic vessel segmentation results in the HVS-External, with red indicating the segmentation and green representing the corresponding labels. Improvements are highlighted with yellow arrows.}
\label{fig:HVS_external2}
\end{figure}
\section{Flexible Convolution Block}
\label{sec:convolution}
Diversifying the receptive fields of convolutions is an effective way to adapt models to targets of different sizes~\cite{zhong2022you}. In our study, the sizes of liver vessels are also various, so diverse receptive fields are beneficial in enhancing the model capability. The flexible convolution block we designed is shown in Figure~\ref{fig:flexibleConv}. To avoid the gridding effect of dilated convolution for extracting local details of vessels, our flexible convolution block provides different receptive fields by stacking the convolutions in parallel rather than using dilated convolution. After the feature maps are fed into this block, they are further encoded by parallel stacked convolutions with different receptive fields~\cite{simonyan2014very}. Then a $1\times1\times1$ convolution integrates all features and compresses the channel for output. Flexible convolution blocks are used at the encoder and the shallow query module (F-Conv in (c) of Figure 2 in the manuscript).

\section{Segmentation Fusion in D2SD}
\label{sec:segfusion}
Vessels of varying sizes exhibit distinct feature representations at different scales. Larger vessels can be effectively reflected in multi-scale feature maps. Yet, for smaller vessels, the information loss caused by successive convolutions and pooling tends to impair their feature representation, which is also one of the motivations why skip connections have been introduced. To mitigate this, the D2SD strategy uses low-cost pre-decoders at multiple scales to capture scale-specific information and aggregate multi-scale outputs for final segmentation, as shown in Figure~\ref{fig:d2sd}. It is important to clarify that the D2SD is distinct from the deep supervision, which does not compute loss for each decoded result. Concretely, the pre-decoder further facilitates the alignment and aggregation between features of vesselness and CT across different scales within the ~$\boldsymbol{SQ}_{i}$ through the use of F-Conv, and then, the interpolation is conducted to obtain pre-decoded results that exhibit varying sensitivities to vessel sizes at different scales, which reduces the impact of varying liver vessel sizes. These results are further fused to become the final segmentation, upon which the loss function is calculated.

\section{Refined Hepatic Vessel Labels}
\label{sec:refine}
In this paper, we use a combined liver vessel segmentation dataset called the HVS dataset. It is based on three publicly available datasets, including LiVS~\cite{gao2023laplacian}, MSD8~\cite{simpson2019large}, and 3DIRCADb~\cite{soler20103d}. 532, 440, and 20 cases are available for the three datasets, respectively. The three publicly available datasets have made an impressive contribution to developing hepatic vessel segmentation models. However, 
some slices of the LiVS dataset are insufficiently labeled, and the labels of the test set of MSD8 are unavailable. The mentioned situations are reflected by labeled ratios (defined by the labeled slices divided by the total slice number of a 3D volume) in Table~\ref{table:LabeledRatios}. Thus, to develop our model, we aim to make the best use of the data and refine these hepatic vessel labels. Fortunately, our clinical cooperator, after carefully checking the labels of the MSD8 training set, considered them to be relatively well labeled. Inspired by this, we first trained our model based on the training set of MSD8 and then used it to infer hepatic vessels of the LiVS dataset and the test set of MSD8. Subsequently, the pseudo labels were fused with the raw labels. Fused labels were checked again and manually corrected by a clinician, to serve as the final hepatic vessel labels in the HVS task. From Table~\ref{table:LabeledRatios}, it can be found that the ratio of labeled slices has been significantly improved after our optimization, especially the LiVS dataset. Moreover, more visualization examples are given in Figures~\ref{fig:RefinedLabels_LiVS} and~\ref{fig:RefinedLabels_MSD8}. Due to the cropping of the CT volume by the organizers of the dataset, the presence of some lesions, such as tumors, and the slice thickness, the continuity of the refined vessel labels is not fully ensured. Still, they are significantly improved compared to the original ones. 
~We also compare the baseline performance trained by the original labels and the refined ones. As indicated by the evaluation metrics in Table~\ref{table:LabelsComparison} and the visualization examples in Figure~\ref{fig:originalvsrefined}, the baseline trained by the refined labels performs better in the HVS task.

\section{Ablation studies}
\label{sec:ablation}
Some visualization examples in ablation studies are shown in Figure~\ref{fig:HVS_abaltion}, it can be seen that our D2SD strategy can extract vessels with diverse sizes more effectively compared to the baseline. Moreover, liver vessel segmentation can not benefit from the simple concatenation fusion between the images and corresponding vesselness filtering results. Instead, our SADF fusion module can better utilize the vesselness filtering result to improve the segmentation accuracy. Besides, it can be observed that the GSB further preserves a reasonable continuity of the vessel tree.

\section{Analysis on HVS-External}
\label{sec:hvs}
In the HVS-External, we included cases with various liver diseases, including two cases of fatty liver, four cases of cirrhosis, twelve cases of liver tumors, and three healthy livers. We analyze the results of HVS-External based on the disease stratification, as shown in Table~\ref{table:Further_HVSExternal}. It can be found that the HarmonySeg achieves the highest mean Dice for patients with fatty liver, tumors, and healthy individuals, and mean HDs are competitive compared with other methods. Furthermore, visualization examples are demonstrated in Figure~\ref{fig:HVS_external2}. The results indicate the robustness of HarmonySeg to various liver diseases and the potential to be applied in clinical practices.


\section{Robustness discussion}
We recognize that the reconnection loss may introduce noise. To address this, we observe that the performance gains of our loss functions following this order:  
$\mathcal{L}_{\text{sup-r}} (2.94\%) > \mathcal{L}_{\text{mix}} (2.01\%) > \mathcal{L}_{\text{spatial}} (1.67\%) > \mathcal{L}_{\text{con}} (1.01\%)
$. The first three losses (\(\mathcal{L}_{\text{sup-r}}\), \(\mathcal{L}_{\text{mix}}\), \(\mathcal{L}_{\text{spatial}}\)) are robust and applicable to various scenarios. In contrast, the reconnection loss \(\mathcal{L}_{\text{con}}\) is specifically designed to address missing vessel segments. To enhance its robustness, we employ two strategies:  
(a) We perform skeletonization on the defined reconnect branches, reducing their pixel width to 1, as shown in Eq.(6) of manuscript. Consequently, the loss applied to these pixels remains slight on average.  
(b) If incorrect pixels are mistakenly defined for reconnection, they can be effectively suppressed by the strong regularization from spatial relationships and mix augmentation invariance.  
Thus, we incorporate the reconnection loss as an additional strategy tailored for vessel segmentation tasks.  
\begin{table}[!t]
\centering
\caption{\textbf{Ablations on recall and precision trade-off}: $\mathcal{L}^+$ for growth, $\mathcal{L}^-$ for suppression.}
\resizebox{\linewidth}{!}{
\begin{tabular}{ccccccc}
\toprule
$\mathcal{L}^{+}_\text{r-sup}$& $\mathcal{L}^{+}_\text{con}$&$\mathcal{L}^{-}_\text{spatial}$&$\mathcal{L}^{-}_\text{mix}$ & Recall ($\%$) & Precision ($\%$) & F1-score ($\%$)\\
\midrule
- & - & - & - & 49.14 & \textbf{84.12} & 62.04 \\
\checkmark &  &  &  & 55.57 & 79.12 & 63.09 \\
\checkmark & \checkmark &  &  & \textbf{64.35} & 71.20 & \underline{65.01} \\
\checkmark &  & \checkmark &  & 50.33 & 79.92 & 59.68 \\
\checkmark &  &  & \checkmark & 53.58 & \underline{80.15} & 62.15 \\
\checkmark & \checkmark & \checkmark & \checkmark & \underline{59.93} & 73.89 & \textbf{66.18} \\
\bottomrule
\end{tabular}}
\label{tab:recall}
\end{table}
\label{sec:robust}
\section{Trade-off between precision and recall}
\label{sec:tradeoff}
We also analyze the recall-precision trade-off. The recall rates for different loss combinations on HVS are presented in \Cref{tab:recall}. Recall is enhanced through relaxed supervision (\(\mathcal{L}_{\text{r-sup}}\)) and branch reconnection (\(\mathcal{L}_{\text{con}}\)), while noise is reduced via spatial consistency (\(\mathcal{L}_{\text{spatial}}\)) and mix equivalence (\(\mathcal{L}_{\text{mix}}\)). We achieved the best trade-off and the highest F1-score when combining all loss functions.

{
    \small
    \bibliographystyle{ieeenat_fullname}
    \bibliography{main}
}

%% file: sec/0_abstract.tex
\begin{abstract}
Accurate segmentation of tubular structures in medical images, such as vessels and airway trees, is crucial for computer-aided diagnosis, radiotherapy, and surgical planning. However, significant challenges exist in algorithm design when faced with diverse sizes, complex topologies, and (often) incomplete data annotation of these structures. We address these difficulties by proposing a new tubular structure segmentation framework named HarmonySeg. 
First, we design a deep-to-shallow decoder network featuring flexible convolution blocks with varying receptive fields, which enables the model to effectively adapt to tubular structures of different scales. Second, to highlight potential anatomical regions and improve the recall of small tubular structures, we incorporate vesselness maps as auxiliary information. These maps are aligned with image features through a shallow-and-deep fusion module, which simultaneously eliminates unreasonable candidates to maintain high precision. Finally, we introduce a topology-preserving loss function that leverages contextual and shape priors to balance the growth and suppression of tubular structures, which also allows the model to handle low-quality and incomplete annotations. 
Extensive quantitative experiments are conducted on four public datasets. The results show that our model can accurately segment 2D and 3D tubular structures and outperform existing state-of-the-art methods. External validation on a private dataset also demonstrates good generalizability. 
\end{abstract}

%% file: sec/1_intro.tex
\section{Introduction}
Accurate segmentation of tubular structures in medical images, such as vessels and airway trees, is important for disease diagnosis, radiotherapy, and surgical planning~\cite{lu2019management, lu2003influence}. Manual delineation of the tubular structure is difficult and time-consuming due to its diverse sizes, complex topologies, and sparse distribution, which affect the efficiency of downstream clinical applications. Moreover, the visibility of small tubular structures is often limited in low-contrast medical images. Therefore, an effective automatic tubular structure segmentation approach is highly desired in clinical practices.
\begin{figure}[!t]
\centering
\includegraphics[width=\linewidth]{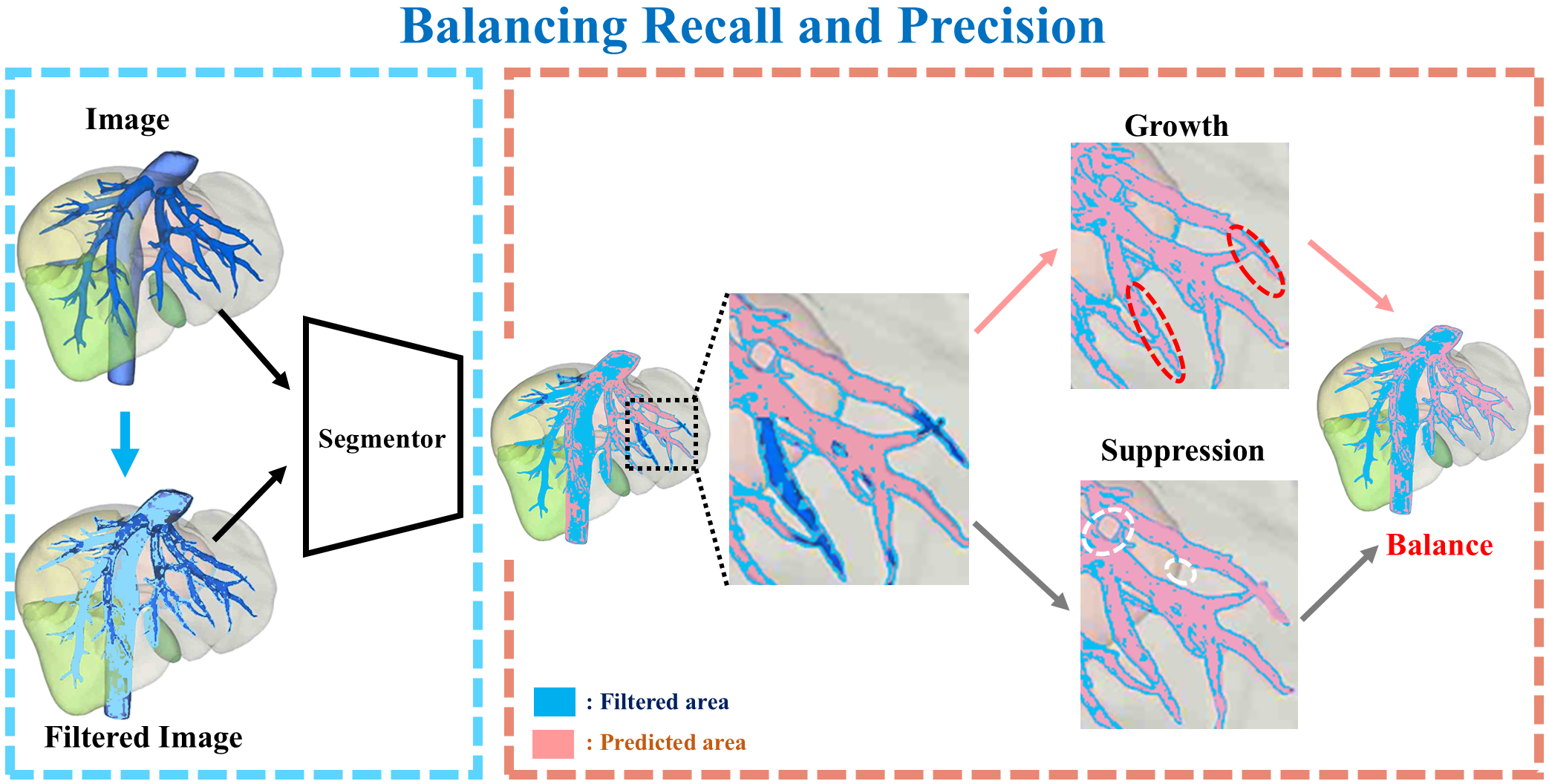}
\caption{
 Motivation of the proposed HarmonySeg. We leverage vesselness to improve recall and filter out outliers based on image features to maintain precision, while applying growth-suppression balanced loss to encourage the growth of vessels at a reasonable noise level in the absence of labels.
 }
\label{fig:intro_fig}
\end{figure}
Various filters were proposed to enhance and segment tubular structures in medical images~\cite{lamy2022benchmark, li2022human}. These filters, producing ``vesselness'' scores, measure how much each pixel resembles a tubular structure. Although effective, they are susceptible to image quality, often struggling with background clutter and noise. As a result, a specially designed post-processing algorithm is necessary, complicating the pipeline and limiting generalizability.
Recent deep learning-based approaches usually
have end-to-end workflows with stronger feature extraction capabilities and higher noise immunity that can obtain accurate tubular structure segmentation without requiring complex post-processing~\cite{hussain2022dilunet, zhong2022you}. However, these methods still have limitations. 
Firstly, 
it is difficult to obtain a large-scale dataset with high-quality labels for model development, as manual annotation of tubular structures is tedious. 
The masks from publicly available datasets~\cite{gao2023laplacian,simpson2019large} are usually sketchy or even missing. Some works have noticed this problem~\cite{wang2020deep, shit2021cldice, araujo2021topological, kirchhoff2024skeleton}, and they typically focused on maintaining the continuity of structures via skeleton growth. However, they did not adequately balance the growth of skeleton and the suppression of noise resulting from overgrowth. 
Secondly, tubular structures vary markedly in sizes and shapes. Pioneering studies have demonstrated that models utilizing multi-scale feature extraction are well-suited for this situation~\cite{hussain2022dilunet, zhong2022you}. 
Some studies introduced specialized convolution blocks for snake-shaped tubular structures, achieving good performance but at high computational cost~\cite{qi2023dynamic}. Others combine traditional techniques, like vesselness filters, with deep learning to improve tubular structure segmentation~\cite{xu2021noisy,garret2024deep,gao2023laplacian,huang2024representing}. However, these models only concatenate the image and filtering results, a shallow integration that overlooks vesselness's potential to enhance deeper feature extraction for small structures.

To accurately segment tubular structures in medical images, we propose a framework named HarmonySeg. As shown in Figure~\ref{fig:intro_fig}, HarmonySeg leverages vesselness filters to enhance recall in two ways. First, we introduce a deep mutual query (DMQ) module that uses cross-attention between the image and vesselness results to boost deep features, especially for small-scale structures. Second, a deep-to-shallow decoding (D2SD) strategy progressively refines segmentation, preserving multi-scale structures. We replace standard convolutions with flexible blocks featuring diverse receptive fields to capture structures of varying sizes. To address partial annotations, we design loss functions to balance structure growth with noise suppression, reducing false positives and compensating for missing labels. Together, these techniques enable HarmonySeg to effectively capture topology and continuity, demonstrating improved recall and precision.
Our contributions are fourfold: 



\textit{1)} We introduce a shallow and deep fusion (SADF) module designed to fully harness the potential of vesselness maps for improving recall while simultaneously ensuring precision by filtering out unwarranted vessel candidates based on image features.


\textit{2)} A deep-to-shallow decoding~(D2SD) strategy is designed to progressively refine the segmentation results with the enhanced features of SADF, which further align and aggregate the features of vesselness and image at different scales, providing varying sensitivities to target sizes and effectively preserve structures.

\textit{3)} We design loss functions that effectively balance tubular structure growth and noise suppression (GSB). These loss functions compensate for missed labels, enhance recall and reduce false positives.

\textit{4)} Extensive experiments carried out on four public datasets validated the performance of HarmonySeg, which can accurately segment 2D and 3D tubular structures, outperforming existing state-of-the-art methods. An external validation on a private dataset also demonstrates its good generalizability. 
\label{sec:intro}

%% file: sec/2_related_work.tex
\section{Related Work}
In this section, we review the existing approaches that involve tubular structure segmentation in medical images.

\noindent\textbf{Vesselness Filtering:}
Vesselness filters can increase the vessels' contrast and suppress the signal of non-vessel structures. They are often used as a preprocessing step for tubular structure segmentation~\cite{lamy2022benchmark,li2022human}. ~\cite{xu2021noisy} and ~\cite{garret2024deep} concatenated images and vesselness filtering results as the model input to highlight the potential vessel regions. Some works also utilized similar strategies to roughly localize potential tubular structures~\cite{huang2024representing,gao2023laplacian}. However, most of these methods simply fused the image and filtered results by concatenating them as input. By contrast, we consider the vesselness filtering result as an independent auxiliary modality, encode it in parallel with the image interactively, and use a novel mutual fusion module to highlight the tubular structure features at different scales.

\noindent\textbf{Feature Extraction and Fusion:}
Flexible convolution with various receptive fields facilitates feature extraction for tubular structures with varying morphology and size~\cite{hussain2022dilunet,zhong2022you,jin2019dunet,dong2022deu}. Most of them introduced deformable convolution with flexible receptive fields and enhanced multi-scale feature fusion modules. For example, Qi et al.~\cite{qi2023dynamic} designed a new dynamic snake convolution to adaptively focus on the slender and tortuous local features of tubular structures. In our framework, we also adopt flexible convolution blocks with diversified receptive fields, incorporated with a multi-scale D2FD strategy, to improve the model's adaptability for tubular structures of various sizes.

\noindent\textbf{Topology Exploration and Preservation:}
Preserving the complex topology is critical for the segmentation of tubular structures. Improving the model architecture is an effective way to achieve this ~\cite{zhang2022progressive,zhao2022graph,li20213d}. Additionally, loss function constraint is also a useful approach to ensure the connectivity of topology. A new centerline Dice (clDice) loss was introduced in~\cite{shit2021cldice} to measure the similarity between the skeleton of prediction and label to guarantee topology connectivity and consistency. Kirchhoff et al.~\cite{kirchhoff2024skeleton} further proposed skeleton recall loss based on the skeleton, which is computationally efficient and suitable for multi-class tubular structure segmentation tasks. However, these methods are applicable only when the labels are complete. In this paper, we introduce a novel loss function to preserve the topology, ensuring the rationality of skeleton growth by mitigating the noise caused by overgrowth, especially in the context of partial labeling.

%% file: sec/3_method.tex
\section{Method}
\begin{figure*}[!t]
\centering
\includegraphics[width=0.8\textwidth]{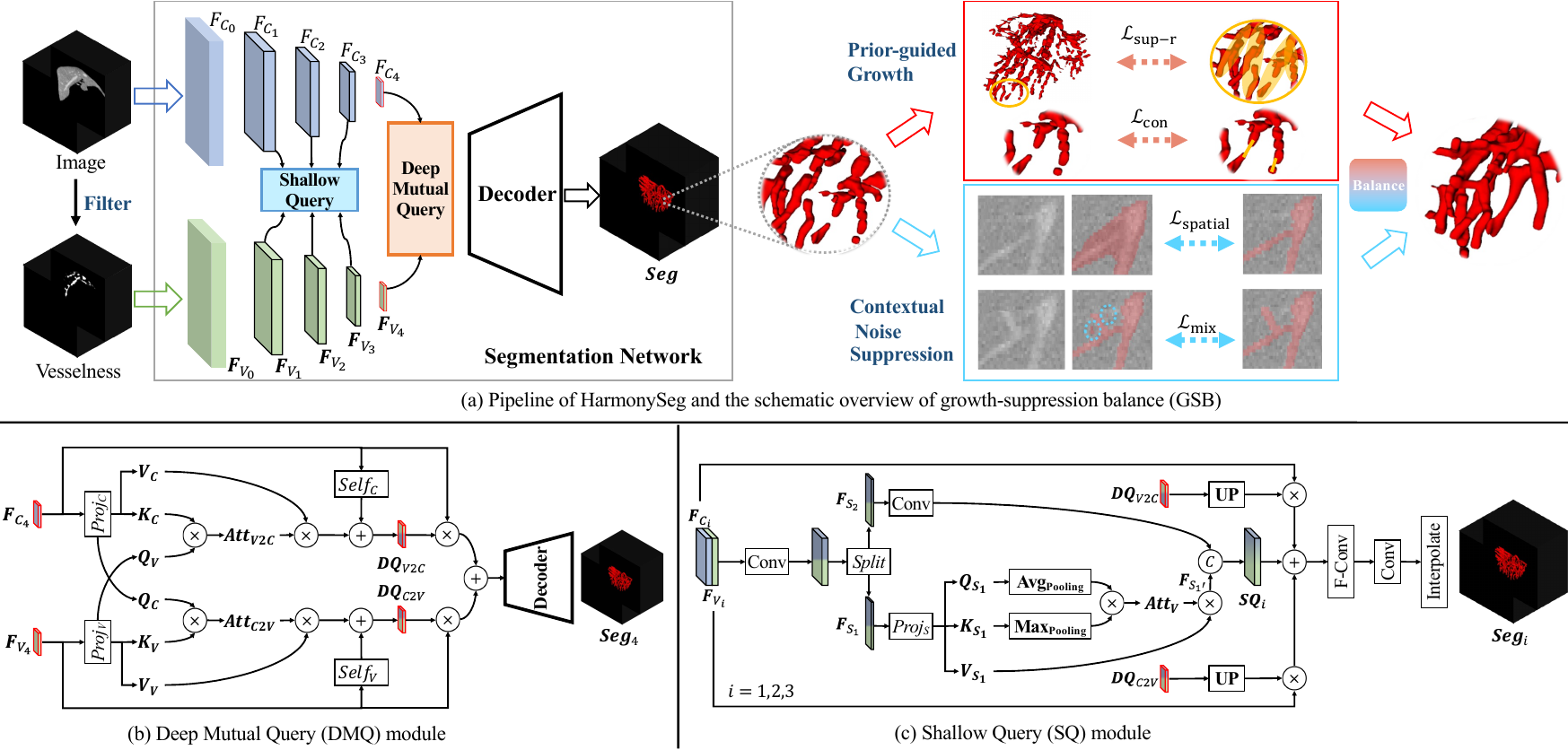}
\caption{The framework of HarmonySeg. Our model takes an image and its filtered results as input, performing multi-level feature fusion to enhance recall. Growth-suppression loss functions are then applied to improve segmentation precision. }
\label{fig:frameworkshow}
\end{figure*}

Our approach is designed for tubular structure segmentation in various medical images. In this section, we take hepatic vessel segmentation in computed tomography (CT) as an example task. The overall architecture is shown in Fig.~\ref{fig:frameworkshow}.

\subsection{Shallow and Deep Fusion~(SADF)}
The vesselness filter is an effective image preprocessing technique that enhances vessel regions while distinguishes vessels from non-vessel structures\footnote{More details are present in Appendix A.}.
We treat the vesselness map as an auxiliary modality and introduce the SADF module to effectively fuse information at both shallow and deep stages. This module comprises two key components: Deep Mutual Query (DMQ) and Shallow Query (SQ).

\noindent\textbf{Deep Mutual Query (DMQ)} is conducted on deep features of both CT and vesselness map, i.e. ${F}_{C_4}$ and ${F}_{V_4}$, which fulfills the integration of global dependencies between the CT and the vesselness and as a basis for decoding. As shown in of of Figure~\ref{fig:frameworkshow}\textbf{(b)}, it can be described as:
\begin{equation}
\boldsymbol{DQ}_{V2C} = Cross(\boldsymbol{Q_V},\boldsymbol{K_C},\boldsymbol{V_C}) + Self(\boldsymbol{F}_{C_4}),
\label{eq:DQv2c}
\end{equation}
where $\boldsymbol{K_C}$ and $\boldsymbol{V_C}$ are the projected key  and value maps from the deep feature $\boldsymbol{F}_{C_4}$ of CT image, and the query maps $\boldsymbol{Q_V}$ is projected from the deep features~$\boldsymbol{F}_{V_4}$ of vesselness map. Next, all of them are injected into the cross-attention mechanism ($Cross$) and further combined with the self-attention ($Self$) results of~$\boldsymbol{F}_{C_4}$ to obtain the enhanced tubular structures' features~$\boldsymbol{DQ}_{V2C}$. In this way, the vesselness highlights the densely vascularized regions of the CT globally, and the global dependence of the CT itself is also protected from being severely affected by noises in the vesselness. Similar processing is done between~$\boldsymbol{F}_{V_4}$ and~$\boldsymbol{F}_{C_4}$ as well to obtain~$\boldsymbol{DQ}_{C2V}$, which aims to mitigate the negative effects of outliers on vesselness by obvious vessels in CT, such as those located at the liver border. 

\noindent\textbf{Shallow Query (SQ)} 
is defined as follows:
\begin{equation}
\boldsymbol{SQ}_{i} = Cat(\boldsymbol{F}_{S_1^{'}}, \boldsymbol{F}_{S_2}),
\end{equation}
\begin{equation}
\boldsymbol{F}_{S_1^{'}}=Self(AvgP(\boldsymbol{Q_{S_1}}), MaxP(\boldsymbol{K_{S_1}}), \boldsymbol{V_{S_1}}),
\label{eq:sq}
\end{equation}
in which~$Cat$,~$AvgP$, and~$MaxP$ refer to concatenation, average, and max pooling, respectively. SQ is implemented on shallow features of CT and vesselness ($\boldsymbol{F}_{C_i}$ and~$\boldsymbol{F}_{V_i}$, $i=1,2,3$). Shallow feature maps contain more accurate spatial information than deep ones, so we utilize vesselness to help locate the potential vessel in the CT image. First, we fuse the shallow features of CT and vesselness, and then equally split them into~$\boldsymbol{F}_{S_1}$ and~$\boldsymbol{F}_{S_2}$ on the channel dimension, to reduce the computational cost of the self-attention mechanism. 
Then,~$\boldsymbol{F}_{S_1}$ is fed into~$Self$, in which pooling operations are used to enhance sensitivity to tiny targets inspired by~\cite{noman2024elgc}. The optimized~$\boldsymbol{F}_{S_1^{'}}$ indicates vessel candidates at a global scale and is further concatenated with~$\boldsymbol{F}_{S_2}$ to recover the original shape. Finally,~$\boldsymbol{SQ}_{i}$ is added with input features improved by the up-sampled ~$\boldsymbol{DQ}_{V2C}$ and~$\boldsymbol{DQ}_{C2V}$ to integrate all vessel information and feed into the parallel multiple decoding stage, see Figure~\ref{fig:frameworkshow}\textbf{(c)}.

\subsection{Deep-to-Shallow Decoding~(D2SD)}
Standard convolution and downsampling operations enrich semantic information while diluting local detail information. 
In our study, the size of the vessel is varied and the topology of vessels is complex, so consistently preserving the spatial information becomes more critical. 
Therefore, we design the D2SD strategy. 
Unlike the common U-Net architecture, HarmonySeg performs decoding progressively at multiple scales from deep to shallow after the complete fusion between CT and vesselness enhanced features achieved by the SADF. Pre-decoding is independently carried out at each shallow scale, taking advantage of the local invariance and detailed spatial information present in the shallow scale features. This allows for the effective localization of vessels that are observable at this scale.
Moreover,~$\boldsymbol{DQ}_{V2C}$ and~$\boldsymbol{DQ}_{C2V}$ are also involved in the pre-decoding after up-sampling, so the fused global dependencies from CT and vesselness further assist in the alignment of the two modality features at the shallow-scale decoding process, which leverages the vesselness to highlight potential vessel regions once again in decoding. Finally, the pre-decoded results~$\boldsymbol{Seg}_{i}$ ($i=1,2,3$) and deep one~$\boldsymbol{Seg}_4$, with varying sensitivities to vessel sizes at different scales, are further fused to get better predictions through a convolution block~\footnote[2]{More details are present in Appendix B and C.}.

Besides, flexible convolution blocks, used in pre-decoding and encoding phases, also benefit the model's ability to extract and aggregate multi-scale features for vessels. We parallelize and stack convolution blocks to provide diverse receptive fields, followed by an~$1\times1\times1$ convolution to adjust the channel number~\footnotemark[2].
~Compared to the direct use of dilated convolution, the convolution stacking approach does not generate a serious grid effect, so it can better preserve local details and be more suitable for vessel segmentation with complex local details.
\begin{figure}[!t]
\centering
\includegraphics[width=0.9\linewidth]{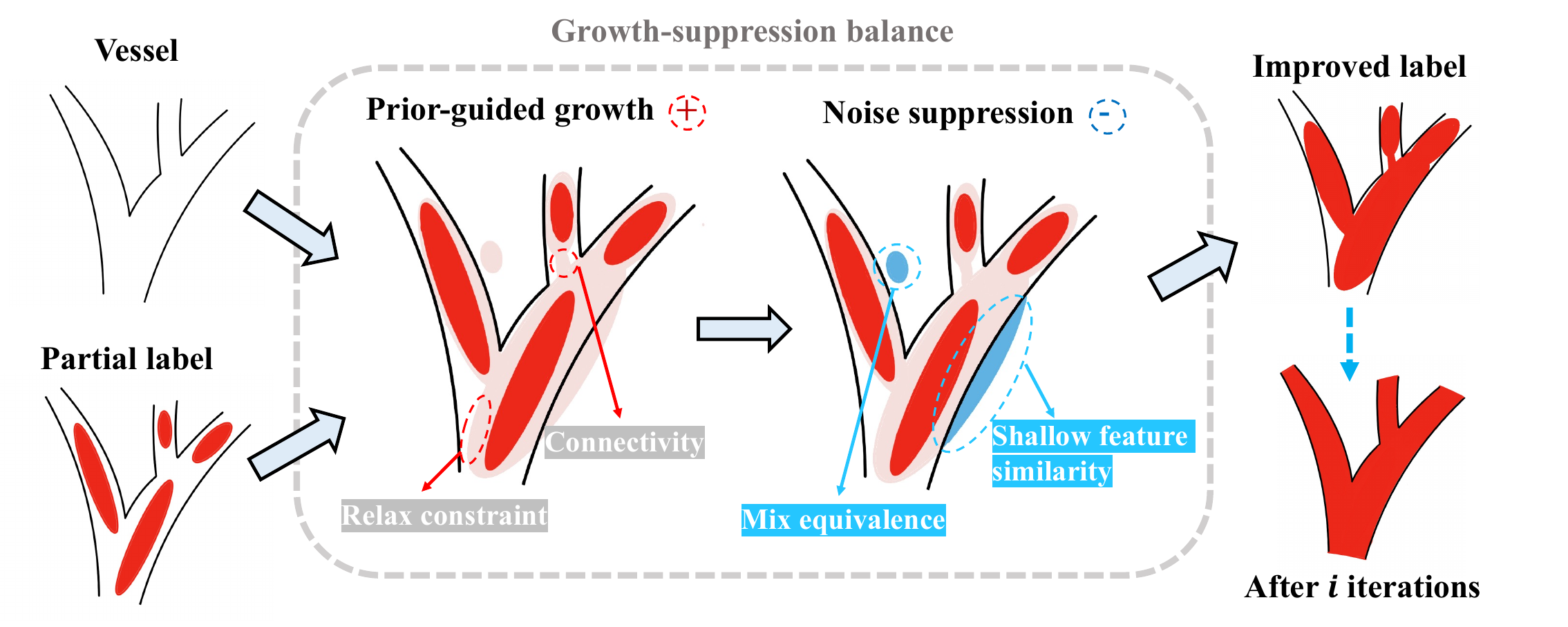}
\caption{Illustration of growth-suppression balance (GSB).}
\label{fig:balance}
\end{figure}

\subsection{Growth-Suppression Balance (GSB)}
In this section, we present the growth-suppression balance strategies for vessel segmentation. As visualized in Figure~\ref{fig:balance}, the process is divided into two stages. 
In the first stage, we utilize local context and shape priors to expand the vessel area, referred to as prior-aware vessel growth. In the second stage, we enforce the mix-equivalence of the predicted segmentation, which helps suppress contextual noise and ensures more refined results. 

\noindent\textbf{Prior-guided vessel growth:} We leverage shape priors to drive the expansion of vessel segmentation. First, due to incomplete annotations, we relax the constraints of the segmentation loss, allowing the segmentation to extend beyond the annotated regions. 
We introduce a region of interest (ROI) term,~$\boldsymbol{R}$, to define a coarse area encompassing all vessels. This region is represented by a bounding box enclosing all annotated vessels. We assume that pixels outside the box belong to the background, which are treated as negative samples. The pixels that belong to annotation $y$ are positive labels. While the remaining pixels represent uncertain vessel regions.
The uncertain-aware prediction $\hat{y}'$ is represented as:
\begin{equation}
\hat{y}' = \underbrace{y\hat{y}}_{\text{positive}} + \beta\underbrace{(y^c \boldsymbol{R}\hat{y})}_{\text{uncertain}} + \underbrace{y^c\boldsymbol{R}^c\hat{y}}_{\text{negative}},
\end{equation}
where $\hat{y}$ is model prediction; the uncertainty ratio $\alpha$ is defined as $\beta = \frac{1}{\log\left(\Sigma y^c/\Sigma y\right)}$; $y^c$ and $\boldsymbol{R}^c$ represent the complement sets of $y$ and $\boldsymbol{R}$, respectively. Then, we derive the relaxed supervison loss $\mathcal{L}_{\text{r-sup}}$ as:
\begin{equation}
\mathcal{L}_{\text{r-sup}} = \mathcal{L}'_{\text{Dice}} + \mathcal{L}_{\text{ce}} = -\frac{y\hat{y}}{y+\hat{y}'}-y\log(y').
\end{equation}

We further reconstruct missing vessel segments, ensuring the completed regions form a continuous, interconnected structure while maintaining density consistency with the surrounding vessels. 
First, we apply the soft-skeletonization method from~\cite{shit2021cldice}, which uses iterative min- and max-pooling operations as proxies for morphological erosion and dilation to capture the vessel’s skeleton results~$\hat{y}^s =f_{\text{skeleton}}(\hat{y})$.
Following~\cite{dulau2023ensuring}, 
We identify the largest connected component (CC) from all smaller ones. For each smaller component, we identify and extract all endpoints. Then, we connect the endpoint of each CC to the nearest endpoint by drawing a line between them.
The resulting paths are represented as $\hat{y}^c = f_{\text{connect}}(\hat{y}^s)$. We then treat these reconnected pixels as pseudo-labels and define a connectivity loss, $\mathcal{L}_{\text{con}}$, to encourage the skeleton outputs to align with their reconnected versions.
\begin{equation}
    \mathcal{L}_{\text{con}} = -\hat{y}^c\log(\hat{y}^s) =  -f_{\text{connect}}(\hat{y}^s)\log(\hat{y}^s),
\end{equation}
where $\mathcal{L}_{\text{con}}$ calculate the cross entropy between the predicted skeleton $\hat{y}^s$ and reconnected skeleton $\hat{y}^c$.

\noindent\textbf{Contextual noise suppression}
We utilize spatial priors and mix-equivalence as guidance to suppress noise in the segmentation results. Firstly, shallow feature similarities are employed to control the growth, ensuring that the expanded regions remain consistent with the original vessel areas regarding density distribution and spatial alignment.
Inspired by~\cite{obukhov2019gated}, we further extend the spatial regularization from 2D images to 3D volumes. We leverage Gaussian kernel $k_{ij}$ to measure the shallow feature similarity between pixels at location $i$ and $j$, \textit{i.e.}, $k_{ij} = exp(-\left[\frac{(l_i-l_j)^2}{2\sigma^2_l}+\frac{(c_i-c_j)^2}{2\sigma^2_c}\right])$.
$l$ and $c$ denote the location and color feature specific to position $i$ and $j$, then we derive the loss of spatial prior with a gated function define the local neighborhood window $\Omega_{r}$ with radius $r$ for each coordinate:
\begin{equation}
\mathcal{L}_{\text{spatial}} = \sum_{i,j \in \Omega_{r}} k_{ij} \hat{y}_i\hat{y}_j
\end{equation}
We further generate mixed samples using the MixUp technique as auxiliary inputs to enforce mix-equivalence. Given two input images, $x_1$ and $x_2$, the mixed image is defined as $x' = \alpha x_1 + (1 - \alpha) x_2$, where $\alpha$ is sampled from uniform distribution. The predicted outputs for these mixed inputs are then constrained by the following loss function:
\begin{equation}
\mathcal{L}_{\text{mix}} = - \frac{\hat{y}' \cdot [\alpha y_1 + (1-\alpha)y_2]}{\|\hat{y}'\|\cdot\|\alpha y_1 + (1-\alpha)y_2\|}
\end{equation}
Finally, we derive the optimization objective ($\mathcal{L}$) that balances the opposing forces of growth and suppression,
\begin{equation}
\mathcal{L} = \underbrace{\mathcal{L}_{\text{r-sup}} + \mathcal{L}_{\text{con}}}_{\text{grow}} +  \underbrace{\lambda(\mathcal{L}_{\text{spatial}} + \mathcal{L}_{\text{mix}})}_{\text{suppress}},
\label{eq:gsb}
\end{equation}
The first two terms promote vessel growth, while the last two focus on noise suppression, and $\lambda$ controls the weight for noise suppression. The proposed loss~$\mathcal{L}$ encourages precise boundary delineation while ensuring effective noise suppression in a well-balanced manner.

%% file: sec/4_experiment.tex
\begin{table*}[]
\centering
\caption{\textbf{Quantitative comparison on HVS task}. The best and second-best results are highlighted in bold and underlined, respectively.}\label{table:HVS}
{
\resizebox{0.75\textwidth}{!}{
\begin{NiceTabular}{ccccccc}
          \CodeBefore
\rowcolor{babyblue!30}{13}
\rowcolor{babyblue!30}{7}
\Body
\midrule
Model                & Dataset &Dice(\%,↑)     & clDice(\%,↑)   & HD(↓)         & ASSD(↓)       & F1-score(\%,↑)\\ \midrule
nnU-Net{~\cite{isensee2021nnu}}            &\multirow{5}{*}{HVS}  & 60.15\textsubscript{±9.49}          & 69.41\textsubscript{±5.43}          & 10.00\textsubscript{±4.22}         & 2.23\textsubscript{±0.85}          & 62.04\textsubscript{±12.15}\\
+ clDice$\mathcal{L}${~\cite{shit2021cldice}}& & 40.82\textsubscript{±11.19}          & 41.98\textsubscript{±10.25}          & 11.78\textsubscript{±4.27}         & 6.81\textsubscript{±2.83}          & 42.22\textsubscript{±10.56}\\
+ SR$\mathcal{L}${~\cite{kirchhoff2024skeleton}}   &  & 61.24\textsubscript{±8.85}          & \underline{71.80}\textsubscript{±4.54}    & 9.94\textsubscript{±4.18}          & \textbf{1.80}\textsubscript{±0.59} & 63.28\textsubscript{±12.43}\\
TransU-Net{~\cite{chen2024transunet}}     &    & 42.49\textsubscript{±17.84}          & 50.11\textsubscript{±18.54}          & 12.96\textsubscript{±7.14}        & 4.33\textsubscript{±3.69}          & 48.54\textsubscript{±21.66}\\
DSC-Net{~\cite{qi2023dynamic}}          &    & \underline{63.57}\textsubscript{±6.83}    & 70.62\textsubscript{±5.57}          & \underline{9.65}\textsubscript{±3.86}    & 2.02\textsubscript{±0.77}           & \underline{65.51}\textsubscript{±13.13}\\
HarmonySeg      &     & \textbf{66.79}\textsubscript{±6.34} & \textbf{72.04}\textsubscript{±5.22} & \textbf{9.60}\textsubscript{±3.98} & \underline{1.96}\textsubscript{±0.73}    & \textbf{67.17}\textsubscript{±14.39}\\ \midrule
nnU-Net{~\cite{isensee2021nnu}}  &     \multirow{5}{*}{HVS-External}        & 63.78\textsubscript{±9.27}          & 69.30\textsubscript{±9.52}          & 3.69\textsubscript{±1.17}          & 2.33\textsubscript{±1.69}          & 64.40\textsubscript{±0.75}\\
+ clDice$\mathcal{L}${~\cite{shit2021cldice}}& & 57.31\textsubscript{±7.23}          & 53.52\textsubscript{±8.20}          & 4.27\textsubscript{±1.13}          & 2.76\textsubscript{±0.99}         & 57.85\textsubscript{±5.98}\\
+ SR$\mathcal{L}${~\cite{kirchhoff2024skeleton}} &    & 71.64\textsubscript{±9.33}          & 78.62\textsubscript{±9.72}          & \underline{3.19}\textsubscript{±1.12}    & \underline{1.32}\textsubscript{±1.04}    & 72.47\textsubscript{±3.13}\\
TransU-Net{~\cite{chen2024transunet}}      &   & 42.33\textsubscript{±22.25}          & 43.63\textsubscript{±21.36}          & 8.03\textsubscript{±7.54}          & 6.55\textsubscript{±9.33}          & 56.10\textsubscript{±30.53}\\
DSC-Net{~\cite{qi2023dynamic}}       &       & \underline{75.83}\textsubscript{±6.19}    & \underline{77.54}\textsubscript{±7.43}    & \textbf{3.15}\textsubscript{±1.16} & \textbf{1.02}\textsubscript{±0.50} & \underline{76.17}\textsubscript{±5.52}\\ 
HarmonySeg      &     & \textbf{77.76}\textsubscript{±6.12} & \textbf{80.61}\textsubscript{±8.07} & 3.55\textsubscript{±1.19}          & 1.40\textsubscript{±1.32}          & \textbf{77.15}\textsubscript{±8.48}\\ \bottomrule
\end{NiceTabular}}}
\end{table*}

\begin{figure}[t]
\centering
\includegraphics[width=0.85\linewidth]{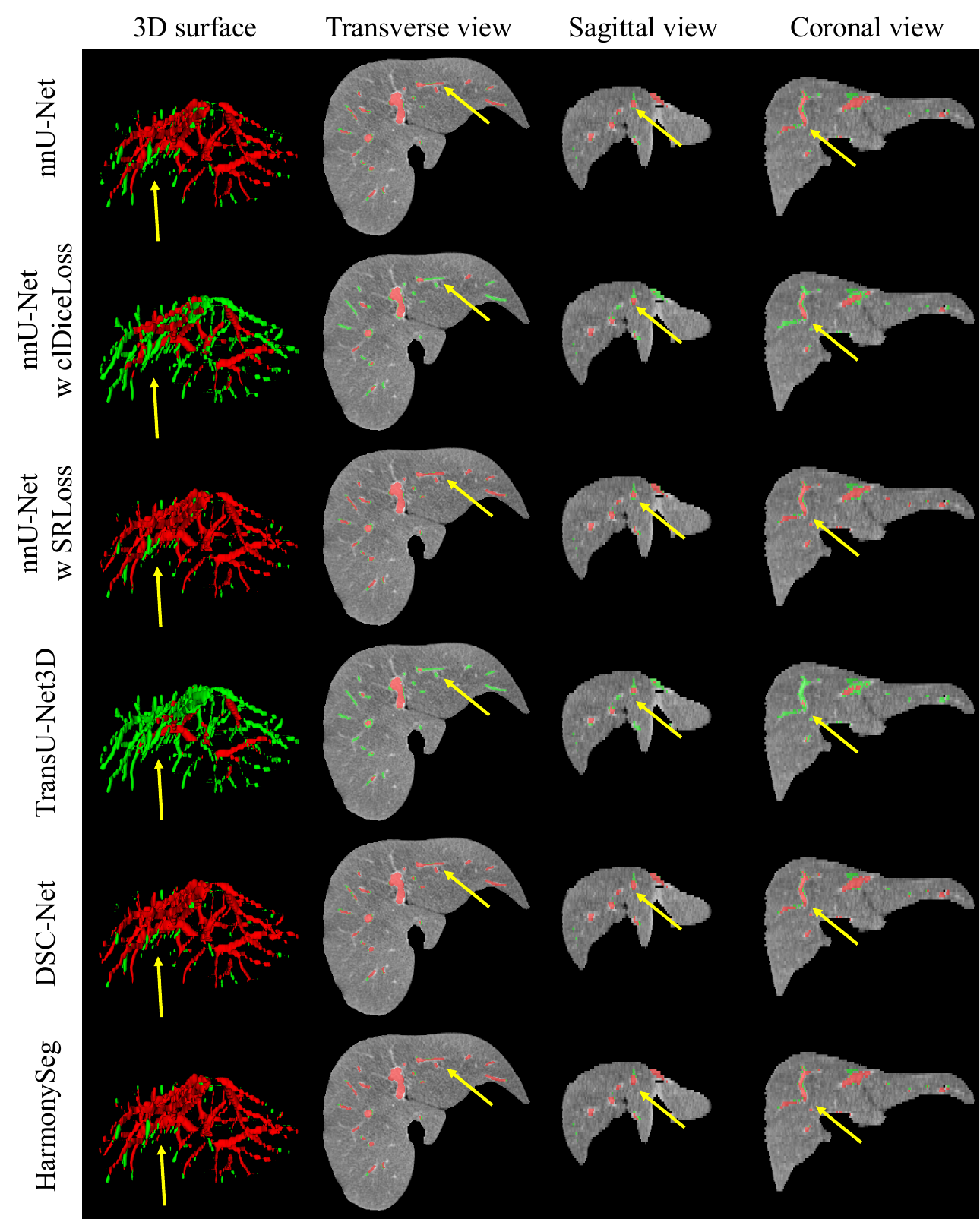}
\caption{Visualization of hepatic vessel segmentation results, with red indicating the segmentation and green representing the corresponding labels. Yellow arrows highlight the improvements.}
\label{fig:HVS_compare}
\end{figure}


\section{Experiments}

\subsection{Dataset}
\textbf{Hepatic vessel segmentation (HVS)}. 
We curated a hybrid hepatic vessel segmentation dataset using contrast enhanced CT scans from 992 patients to evaluate our method. This dataset combines three public datasets: LiVS~\cite{gao2023laplacian} (532 volumes), MSD8~\cite{simpson2019large} (440 volumes), and 3DIRCADb~\cite{soler20103d} (20 volumes). LiVS has incomplete labels, and MSD8's test set labels are not publicly available. These label gaps hinder topology capture and loss computation for correctly segmented regions without labels. To fix this, we first trained a model using MSD8's training set since it has relatively complete labels, and then used it to refine labels on LiVS and MSD8’s test set, followed by manual refinement by experts~\footnote[3]{Details are displayed in the supplementary materials.}. 
\noindent We applied Hounsfiled Unit (HU) value clipping (-100, 200) and liver cropping to the MSD8 and 3DIRCADb volumes, following the preprocessing steps for LiVS. Finally, volumes with refined labels from LiVS and MSD8 were used for training and validation, while 3DIRCADb volumes were used for testing.

\noindent\textbf{HVS-External}. An external testing set was collected to further assess HarmonySeg's effectiveness, which contains 21 cases from our collaborating hospital with hepatic vessel labels manually delineated by experienced experts and underwent the same preprocessing as HVS. It was tested directly using the models trained on HVS.

\noindent\textbf{Retinal vessel segmentation (RVS)}. The DRIVE dataset~\cite{staal2004ridge} consisted of 40 2D digital retinal images. They were equally divided into 20 images for training and 20 for testing. The retinal vessel labels were manually delineated by ophthalmological experts.

\noindent\textbf{Airway tree segmentation (ATS)}. The airway tree segmentation dataset~\cite{qin2019airwaynet, wang2024accurate} includes 90 chest CT volumes, in which 70 and 20 were collected from the LIDC dataset~\cite{armato2011lung} and the training set of EXACT’09 dataset ~\cite{lo2012extraction}, respectively. These volumes were further split into 50, 20 and 20 volumes for training, validation and testing, respectively. Similar to the HVS task, the HU clip (-1350, 150) was performed for the volumes in the ATS task.

\noindent\textbf{Coronary artery segmentation (CAS)}. The coronary artery segmentation dataset was established based on the ImageCAS dataset~\cite{zeng2023imagecas}, which captured data from 1000 patients. The dataset was divided into 700, 50, and 250 cases for training, validation, and testing, respectively. An HU clip (-400, 500) was also conducted to optimize the observability of coronary arteries.


\begin{table}[]
\centering
\caption{\textbf{Dice(\%) for different vessel sizes on the HVS task.} Vessel branches were categorized into small ($<$5mm), medium (5$-$10mm), and large ($\geq$10mm) based on diameter.}
\label{table:HVS_differentsize}
\resizebox{0.85\linewidth}{!}{
\begin{NiceTabular}{cccc}
          \CodeBefore
\rowcolor{babyblue!30}{7}
\Body
\toprule
Model                & Small     & Medium & Large  \\ 
\hline
nnU-Net{~\cite{isensee2021nnu}}              & 24.04\textsubscript{±14.77} & 39.02\textsubscript{±18.04} & 54.11\textsubscript{±15.77}  \\
+ clDice$\mathcal{L}${~\cite{shit2021cldice}} & 8.08\textsubscript{±8.27} & 27.93\textsubscript{±16.88} & 29.51\textsubscript{±15.65}   \\
+ SR$\mathcal{L}${~\cite{kirchhoff2024skeleton}}     & 25.02\textsubscript{±10.93} & 
40.16\textsubscript{±16.76} & 52.26\textsubscript{±15.47} \\
TransU-Net{~\cite{chen2024transunet}}         & 11.07\textsubscript{±9.35} & 25.45\textsubscript{±16.21} & 35.15\textsubscript{±20.93}  \\
DSC-Net{~\cite{qi2023dynamic}}               & \underline{25.66}\textsubscript{±8.93} & \underline{40.85}\textsubscript{±8.99}  & \underline{57.64}\textsubscript{±11.30}  \\
HarmonySeg           & \textbf{27.93}\textsubscript{±13.59} & \textbf{40.97}\textsubscript{±7.47}  & \textbf{58.46}\textsubscript{±12.87} \\  \bottomrule
\end{NiceTabular}}
\end{table}

\begin{table}[]
\centering
\caption{\textbf{Quantitative comparison on the RVS task.} 
}\label{table:RVS}
\resizebox{0.85\linewidth}{!}{
\begin{NiceTabular}{cccc}
          \CodeBefore
\rowcolor{babyblue!30}{8}
\Body
\toprule
Model & Dice(\%,↑) & clDice(\%,↑) & HD(↓) \\ \midrule
U-Net{~\cite{ronneberger2015u}} & 80.73\textsubscript{±1.77} & 79.66\textsubscript{±4.00} & 6.86\textsubscript{±0.56} \\
TransU-Net{~\cite{chen2021transunet}} & 80.56\textsubscript{±2.14} & 79.02\textsubscript{±5.05} & 6.83\textsubscript{±0.52} \\
CS\textsuperscript{2}-Net{~\cite{mou2021cs2}} & 77.53\textsubscript{±2.94} & 74.88\textsubscript{±5.27} & 6.90\textsubscript{±0.48} \\
DCU-Net{~\cite{yang2022dcu}} & 80.83\textsubscript{±1.99} & 80.19\textsubscript{±4.80} & \underline{6.68}\textsubscript{±0.49} \\
DSC-Net{~\cite{qi2023dynamic}} & \textbf{81.85}\textsubscript{±1.74} & \underline{81.16}\textsubscript{±4.54} & \underline{6.68}\textsubscript{±0.49} \\ 
nnU-Net{~\cite{isensee2021nnu}} & 80.13\textsubscript{±1.60} & 78.82\textsubscript{±3.77} & 7.61\textsubscript{±0.47}\\ 
HarmonySeg & \underline{81.33}\textsubscript{±1.61} & \textbf{82.03}\textsubscript{±3.58} &  \textbf{6.51}\textsubscript{±0.83} \\ \bottomrule
\end{NiceTabular}}
\end{table}


\subsection{Implementation Details}
We implemented HarmonySeg in PyTorch and ran experiments on an NVIDIA A100 GPU with 80 GB of memory. The training was performed using the Adam optimizer with an initial learning rate of 1e-4 and a polynomial decay strategy. The preprocessing followed the same scheme as the nnU-Net~\cite{isensee2021nnu}. For the vesselness filter, we follow parameters from ~\cite{lamy2022benchmark}. The suppression weight $\lambda$ is empirically set to 1, with its impact validated in the ablation study.

\noindent\textbf{Metrics:} We used a series of quantitative evaluation metrics based on overlap, distance, and connectivity to comprehensively measure the performance of the model. Concretely, the HVS task employed pixel-wise 
Dice, centerline Dice (clDice), Hausdorff distance (HD), average symmetric surface distance (ASSD) and F1-score
as evaluation metrics~\cite{gao2023laplacian}. Similarly, the RVA task also used Dice, clDice, and HD~\cite{qi2023dynamic}. The ATS task focused on connectivity, so branch detected (BD), tree length detected (TLD), and precision, were used as evaluation metrics~\cite{lo2012extraction}. The CAS task introduced Dice, HD, and average Hausdorff distance (AHD) to assess the model performance~\cite{zeng2023imagecas}. In addition, we thoroughly compared with various benchmarks published on the dataset of RVS, ATS, and CAS.

\begin{table}[!t]
\centering
\caption{\textbf{Results on ATS task}. Prec denotes Precision.}\label{table:ATS}
\resizebox{0.85\linewidth}{!}{
\begin{NiceTabular}{cccc}
\CodeBefore
\rowcolor{babyblue!30}{8}
\Body
\toprule
Model & BD(\%,↑) & TLD(\%,↑) & Prec(\%,↑) \\ \midrule
Juarez et al.{~\cite{garcia2019joint}} & 69.2\textsubscript{±25.4} & 53.5\textsubscript{±20.9} & \textbf{99.9}\textsubscript{±0.1} \\
WingsNet{~\cite{zheng2021alleviating}} & 89.2\textsubscript{±5.8} & 77.1\textsubscript{±5.7} & 99.0\textsubscript{±0.8} \\
CFDA{~\cite{zhang2022cfda}} & 90.9\textsubscript{±6.7} & 78.9\textsubscript{±8.1} & \underline{99.1}\textsubscript{±0.6} \\
Qin et al.{~\cite{qin2021learning}} & 90.9\textsubscript{±8.8} & 80.7\textsubscript{±9.9} & 98.4\textsubscript{±1.0} \\
Zheng et al.{~\cite{zheng2021refined}} & \underline{91.1}\textsubscript{±5.5} & 80.1\textsubscript{±6.6} & 98.9\textsubscript{±0.7} \\ 
nnU-Net{~\cite{isensee2021nnu}} & 90.0\textsubscript{±30.0} & \underline{91.0}\textsubscript{±5.0} & 88.7\textsubscript{±5.7} \\ 
HarmonySeg & \textbf{95.0}\textsubscript{±21.8} & \textbf{92.3}\textsubscript{±4.5} & 91.1\textsubscript{±2.9} \\ \bottomrule
\end{NiceTabular}}
\end{table}

\begin{table}[]
\centering
\caption{\textbf{Quantitative comparison on the CAS task}.}\label{table:CAS}
\resizebox{0.85\linewidth}{!}{
\begin{NiceTabular}{cccc}
          \CodeBefore
\rowcolor{babyblue!30}{8}
\Body
\toprule
Model & Dice(\%,↑) & HD(↓) & AHD(↓) \\ \midrule
Shen et al.{~\cite{8675935}} & 80.58 & 28.67 & \underline{0.85} \\
Chen et al.{~\cite{chen2019coronary}} & 72.01 & 40.96 & 3.07 \\
Kong et al.{~\cite{kong2020learning}} & 68.78 & 30.34 & 1.43 \\
Wolterink et al.{~\cite{wolterink2019graph}} & 70.61 & 27.87 & 1.24 \\
U-Net++{~\cite{zhou2018unet++}} & \underline{82.96} & \underline{27.22} & \bf 0.82 \\ 
nnU-Net{~\cite{isensee2021nnu}} & 75.46 & 34.36 & 0.91 \\ 
HarmonySeg & \textbf{83.24} & \textbf{26.42} & 0.88 \\ \bottomrule
\end{NiceTabular}}
\end{table}

\noindent\textbf{Baselines:} We compare HarmonySeg with three state-of-the-art approaches designed for tubular structure segmentation in medical images and two widely-used general medical segmentation models, including nnU-Net~\cite{isensee2021nnu}, nnU-Net with clDice loss (clDice$\mathcal{L}$~\cite{shit2021cldice}), nnU-Net with skeleton recall loss (SR$\mathcal{L}$~\cite{kirchhoff2024skeleton}), TransU-Net~\cite{chen2024transunet} and DSC-Net~\cite{qi2023dynamic}. nnU-Net is a representative baseline in medical image segmentation, and its cooperation with the clDice and the skeleton recall loss are two effective improved variants for tubular structure segmentation. TransU-Net also performed well in medical image segmentation by integrating the local invariance of convolution and global dependencies of the transformer. DSC-Net utilized a dynamic snake convolution to capture more topological information so that precise vessel segmentation was achieved. The comparison experiment was conducted on all three datasets. It should be noted that our model evaluated on RVS did not involve GSB because the retinal vessel is so well labeled that no further growth is required. Our models evaluated on ATS and CAS did not involve SADF because these two datasets 
are not suitable for the vesselness filter as adjacent structures can cause serious interference.
\begin{figure}[!t]
\centering
\includegraphics[width=\linewidth]{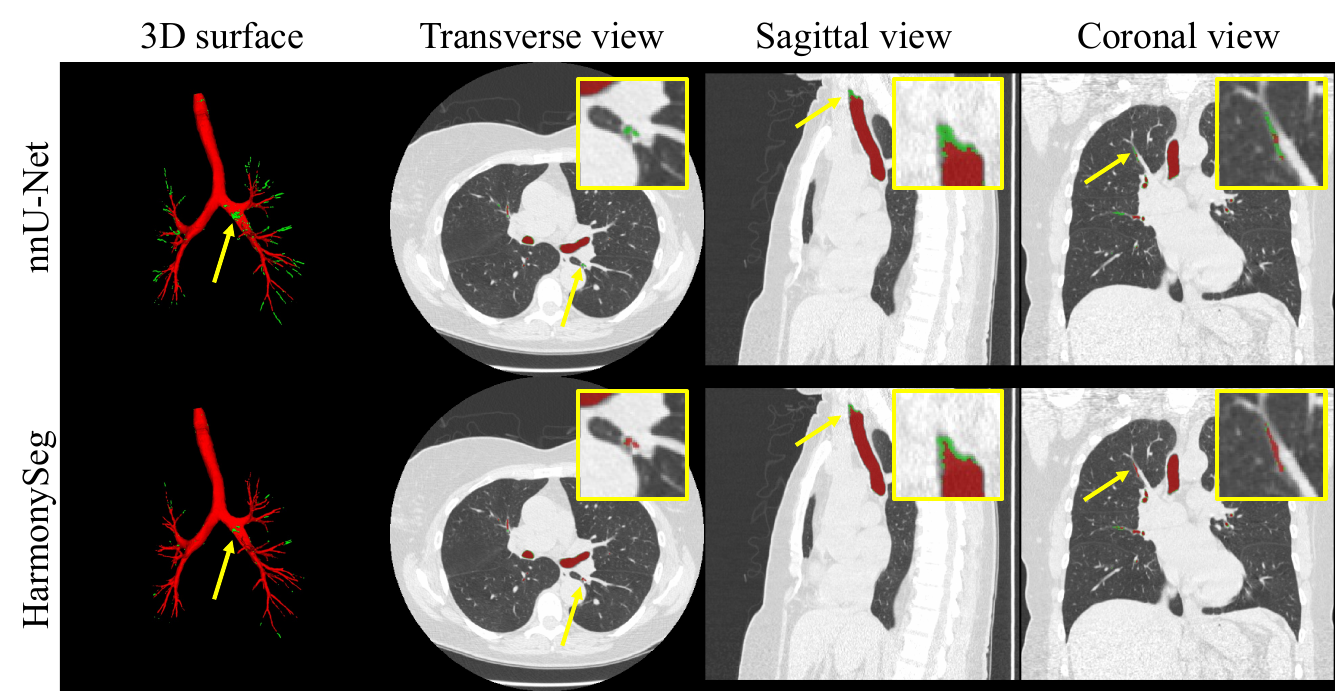}
\caption{Qualitative comparison of airway tree segmentation.
}
\label{fig:ATS_visual}
\end{figure}

\begin{figure}[!t]
\centering
\includegraphics[width=\linewidth]{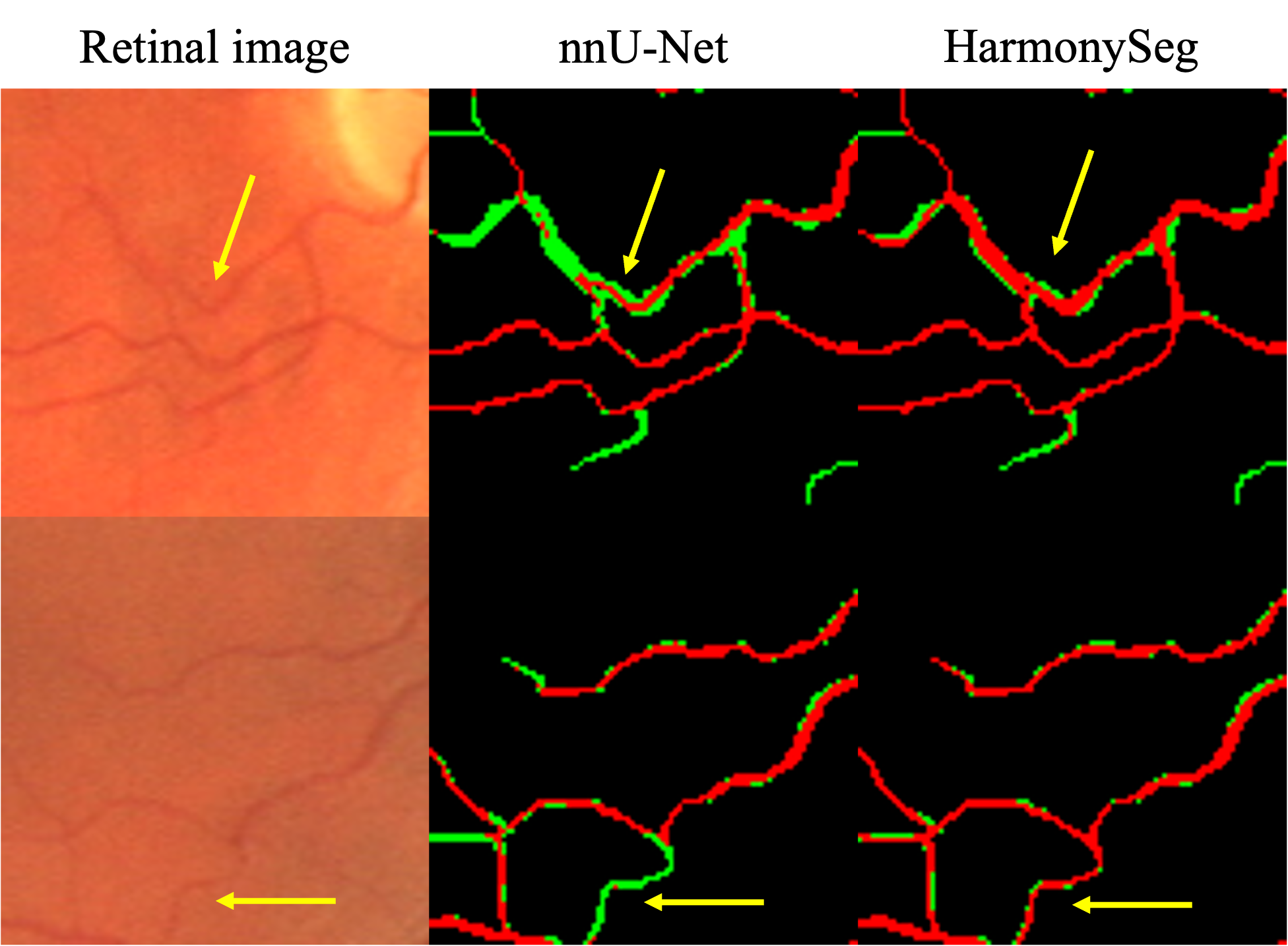}
\caption{Qualitative comparison on retinal vessel segmentation.}
\label{fig:RVS_visual}
\end{figure}

\begin{table*}[]
\centering
\caption{\textbf{Ablation study of HarmonySeg components on the HVS task}. \#2 denotes the model using the simple concatenation of CT and vesselness maps as the input, denoted by \textbf{VS\textsubscript{Cat}}.
}\label{table:HVS_ablation}{
\resizebox{0.88\linewidth}{!}{
\begin{NiceTabular}{ccccccccc}
\CodeBefore
\rowcolor{babyblue!30}{7}
\Body
\toprule
Methods&D2SD & SADF & GSB &  Dice(\%,↑) & clDice(\%,↑)& HD(↓) & ASSD(↓) &F1-score(\%,↑)\\ \midrule
\#1&- & - & - &  60.15\textsubscript{±9.49} & 69.41\textsubscript{±5.43} & 10.00\textsubscript{±4.22} & 2.23\textsubscript{±0.85} &62.04\textsubscript{±12.15}\\
\#2&- & \textbf{VS\textsubscript{Cat}} & -  & 60.07\textsubscript{±8.78} & 68.63\textsubscript{±5.23}& 10.05\textsubscript{±4.21} &2.32\textsubscript{±0.83} &61.82\textsubscript{±11.74}\\
\#3 & \CheckmarkBold &  &  & 63.23\textsubscript{±6.74} & 70.30\textsubscript{±5.79}& \underline{9.74}\textsubscript{±3.76} & 2.27\textsubscript{±0.89} &65.52\textsubscript{±13.97}\\
\#4 & \CheckmarkBold & \CheckmarkBold &  & \underline{64.20}\textsubscript{±6.53} & \textbf{73.08}\textsubscript{±5.06} & 10.22\textsubscript{±3.91} & \textbf{1.79}\textsubscript{±0.62} &\underline{66.43}\textsubscript{±13.92}\\
\#5 &  &  & \CheckmarkBold &64.00\textsubscript{±6.10} & 71.65\textsubscript{±4.91} & 9.90\textsubscript{±3.97} &\underline{1.81}\textsubscript{±0.61} & 66.18\textsubscript{±13.56}\\
\#6 & \multicolumn{3}{c}{HarmonySeg} & \textbf{66.79}\textsubscript{±6.34} & \underline{72.04}\textsubscript{±5.22} & \textbf{9.60}\textsubscript{±3.98} &1.96\textsubscript{±0.73} &\textbf{67.17}\textsubscript{±14.39}\\ \bottomrule
\end{NiceTabular}}}
\end{table*}
\subsection{Performance Comparison}

\textbf{Hepatic vessel segmentation}: Quantitative evaluation metrics are summarized in Table~\ref{table:HVS}. Some visualization examples are given in Figure~\ref{fig:HVS_compare}. As indicated in Table~\ref{table:HVS}, HarmonySeg achieves the best performance on both of the two commonly used segmentation evaluation metrics, Dice and HD, with improvements of 5.1\% and 0.5\% compared to those of the second best model, respectively. Moreover, our model also performs competitively in the comparison of other metrics. Taking F1-score as an example, our model maintains an improved balance between precision and recall, and does not have an obvious drop-off between them like other models. In addition, the Dice comparison in Table~\ref{table:HVS_differentsize} further reveals the adaptability of our model for different vessel sizes, 
especially the better accuracy in small vessels. Observing segmentation results in Figure~\ref{fig:HVS_compare}, it can be found that our 3D hepatic vessel tree is more complete compared with others. In 2D views, our segmentation is also accurate and has good continuity for both large vessels and tiny branches.

\noindent\textbf{Results on HVS-External}: 
The evaluation metrics of HVS-External in Table~\ref{table:HVS} further reveal HarmonySeg's generalizability on external data. It achieves the best mean Dice and clDice, with 2.5\% and 4.0\% improvements compared with the second. Other metrics are also competitive. This generalizability indicates that our model has potential for application in clinical practice.

\noindent\textbf{Retinal vessel segmentation}: Table~\ref{table:RVS} shows the superiority of our framework even without the GSB. The highest clDice is achieved by HarmonySeg, which has an increment of 1.1\% in comparison with the second. This improvement is also visualized in Figure~\ref{fig:RVS_visual}. As highlighted in the figure, our model preserves continuity for tiny branches effectively. Further, the HD of HarmonySeg also decreases by 2.5\%.

\noindent\textbf{Airway tree segmentation}: Our model reports the highest BD and TLD with competitive precision in Table~\ref{table:ATS} for airway tree segmentation. Compared with those of the second-best model, the mean BD and TLD raise 4.3\% and 1.4\%, respectively. The visualization in Figure~\ref{fig:ATS_visual} also indicates that more airway tree branches are extracted by our model.

\begin{figure}[!t]
\centering
\includegraphics[width=\linewidth]{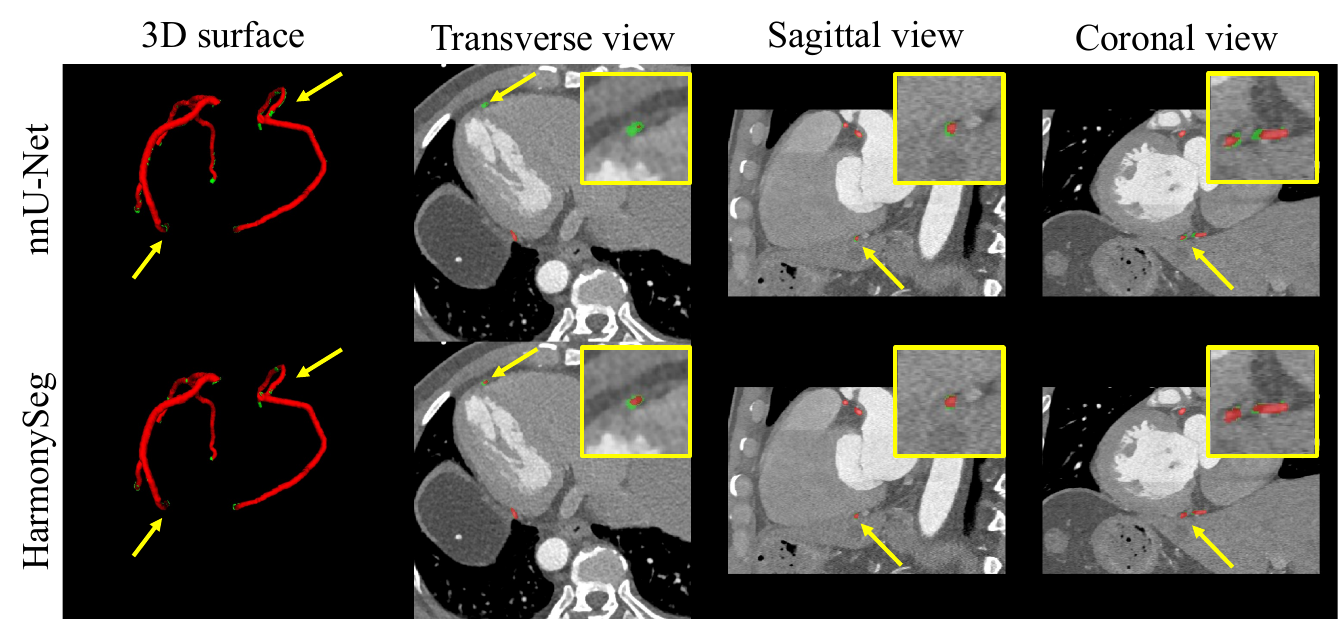}
\caption{Qualitative comparison of coronary artery segmentation.}
\label{fig:CAS_visual}
\end{figure}

\noindent\textbf{Coronary artery segmentation}: The effectiveness of HarmonySeg in coronary artery segmentation is demonstrated with the best Dice and HD in Table~\ref{table:CAS}. Moreover, it can be observed in the visualization example of Figure~\ref{fig:CAS_visual} that HarmonySeg obtains a more complete coronary artery in 3D views and captures more tiny vessels in 2D views.

\begin{figure}[!t]
\centering
\includegraphics[width=\linewidth]{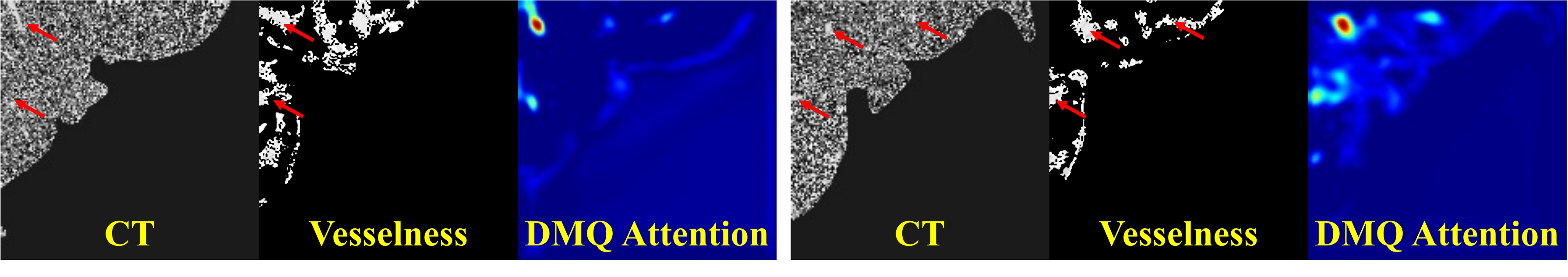}
\caption{Attention visualization examples of the DMQ module.}
\label{fig:heatmap}
\end{figure}

\subsection{Ablation studies}
We conduct four validation studies on the HVS dataset. First, we demonstrate the effectiveness of HarmonySeg's three key components: D2SD (Deep-to-Shallow Decoding), SADF (Shallow and Deep Fusion), and GSB (Growth-Suppression Balance). Next, we perform ablation studies on the GSB module to evaluate the impact of different loss combinations.


\noindent\textbf{Ablations of HarmonySeg's key components:}  The results are summarized in \Cref{table:HVS_ablation}. 
Using nnUNet as the backbone, the model enhanced with D2SD (Deep-to-Shallow Decoding, Model \#3) achieves significant improvements, with a 5.1\% increase in Dice and a 2.6\% reduction in HD, by leveraging local invariance and detailed spatial information from shallow-scale features. 
Regarding vesselness utilization, a comparison between Model \#1 and Model \#2 reveals that simply concatenating CT and vesselness maps (\textbf{VS\textsubscript{Cat}}) does not yield notable performance gains, indicating that this fusion method fails to effectively enhance potential vessel regions.
To better exploit vesselness information, we introduce SADF (Shallow and Deep Fusion) in Model \#4. The results show improvements in clDice and ASSD, demonstrating that SADF, combined with D2SD, more effectively utilizes vesselness to highlight vessel regions. 
Through deep querying with CT, the model successfully identifies and focuses on actual vessels. Attention visualization examples are shown in Figure~\ref{fig:heatmap}.
By comparing model \#1 and model \#5 with GSB in Table~\ref{table:HVS_ablation}, GSB improves the Dice score for nnUNet by 6.4\%, highlighting its effectiveness in extracting more branches precisely. Visual comparisons are provided in Appendix E.
\begin{table}[!t]
	\caption{
    {\textbf{Effectiveness of loss functions}.}
    }\label{tab:ablation_loss}
    \vspace{-.1in}
	\centering
		\resizebox{\linewidth}{!}{
			\begin{NiceTabular}{cccc|cc}
				\toprule
				\multicolumn{4}{c}{Loss functions}&\multicolumn{2}{c}{HVS}\\
				\cmidrule(lr){1-4}\cmidrule(lr){5-6}
				$\mathcal{L}_{\text{r-sup}}$&$\mathcal{L}_{\text{con}}$&$\mathcal{L}_{\text{spatial}}$&$\mathcal{L}_{\text{mix}}$&\multicolumn{1}{c}{Dice(\%,↑)}&\multicolumn{1}{c}{ASSD(↓)}\\
                \midrule
                -&-&-&-&60.15$_{\pm9.49}$&2.23$_{\pm0.85}$\\
                \CheckmarkBold&&&&\underline{63.09}$_{\pm7.22}$&1.89$_{\pm0.70}$\\
                &\CheckmarkBold&&&61.26$_{\pm11.1}$&1.76$_{\pm0.57}$\\
                &&\CheckmarkBold&&61.82$_{\pm11.6}$&\textbf{1.71}$_{\pm0.59}$\\
                &&&\CheckmarkBold&62.16$_{\pm7.56}$&\underline{1.74}$_{\pm0.57}$\\ \CheckmarkBold&\CheckmarkBold&\CheckmarkBold&\CheckmarkBold&\textbf{64.00}$_{\pm6.10}$&1.81$_{\pm0.61}$\\
                \bottomrule
		\end{NiceTabular}}
        \vspace{-.1in}
\end{table}

\noindent\textbf{Ablatons of loss functions:}
We evaluate the GSB module by analyzing noise combinations, parameter sensitivity, and computational cost. The robustness of the loss functions and the trade-off between recall and precision are discussed in detail in Appendix G and Appendix H of the supplementary material, respectively. Table~\ref{tab:ablation_loss} show the combination of loss functions: $\mathcal{L}_{\text{r-sup}}$, $\mathcal{L}_{\text{con}}$, $\mathcal{L}_{\text{spatial}}$, and $\mathcal{L}_\text{mix}$ individually enhance performance by 2.94\%, 1.11\%, 1.67\%, and 2.01\%, respectively. When combined, they synergistically increase the average Dice from 60.15\% to 64.0\%. 
We then investigate the impact of the noise suppression weight \(\lambda\) in the combined loss function, 
as in Eq.~\eqref{eq:gsb}.
As \Cref{tab:lambda} shows, Dice improves as \(\lambda\) increases from 0 to 1, but experiences a slight decline when \(\lambda\) exceeds 1. The combined loss computation time ranges from 0.96s to 4.27s per batch, averaging 2.43s. To optimize efficiency, the reconnection loss can be activated after a warm-up phase using other loss functions.

\begin{table}[!t]
\centering
\caption{\textbf{Parameter sensitivity} on the suppression loss weight $\lambda$.}
\resizebox{\linewidth}{!}{
\begin{tabular}{lcccccc}
\toprule
$\lambda$ & 0 & 0.5&0.75&1&1.5&2\\
Dice ($\%$)& 61.82 & 62.14& 63.45& 64.00 &63.90 &63.64\\
\bottomrule
\end{tabular}}
\label{tab:lambda}
\end{table}


%% file: sec/5_conclusion.tex
\section{Conclusion}
In this paper, we propose HarmonySeg for tubular structure segmentation in medical images. Our model used the deep-to-shallow decoding strategy
to enhance the model's adaptability to tubular structures of different sizes. The shallow query and deep mutual query fusion between input images and vesselness filtering results can highlight the potential regions where tubular structures exist. Moreover, we design loss functions to achieve a balance between vessel growth and noise suppression, compensating for the supervision with missing labels. Our model was comprehensively evaluated on four publicly available datasets and the results consistently demonstrated its superiority. A potential improvement is to further integrate the vesselness filter into the network through convolutional operations, combining it with the CT input to form a truly unified entity.